\documentclass[reprint,amsmath,amssymb,aps,prd]{revtex4-2}

\usepackage{graphicx}
\usepackage{dcolumn}
\usepackage{array}
\usepackage[T1]{fontenc} 
\usepackage{titlesec}
\usepackage[normalem]{ulem}
\usepackage{color}
\usepackage{bm}
\usepackage{multirow}
\usepackage{tikz}
\usepackage{mathtools}

\usepackage{cellspace,tabularx}
\usepackage{comment}
\usepackage{xcolor}
\usepackage[dvipsnames]{xcolor}
\usepackage{xcolor}
\usepackage{subfigure}  
\newcommand{\red}[1]{\textcolor{red}{#1}}
\newcommand{\orange}[1]{\textcolor{orange}{#1}}
\newcommand{\cyan}[1]{\textcolor{cyan}{#1}}

\def\nn{\nonumber}
\def\ds{{\rm d}s}
\def\da{{\rm d}a}

\def\R2{{\mathcal R}^2}

\def\btc{\begin{tcolorbox}}
\def\etc{\end{tcolorbox}}
\def\be{\begin{equation}}
\def\ee{\end{equation}}
\def\barr{\begin{array}{lr}}
\def\earr{\end{array}}
\def\bea{\begin{eqnarray}}
\def\eea{\end{eqnarray}}

\def\nn{\nonumber}

\def\a{\alpha}

\def\e{\epsilon}

\def\k{\kappa}

\def\o{\omega}

\def\s{\sigma}

\def\D{\Delta}

\def\L{\Lambda}

\def\S{\Sigma}

\def\ha{\frac{1}{2}}

\def\nn{\nonumber}
\def\ds{{\rm d}s}

\def\da{{\rm d}a}

\def\be{\begin{equation}}
\def\ee{\end{equation}}
\def\barr{\begin{array}{lr}}
\def\earr{\end{array}}
\def\bea{\begin{eqnarray}}
\def\eea{\end{eqnarray}}
\def\nn{\nonumber}

\def\a{\alpha}

\def\e{\epsilon}

\def\k{\kappa}

\def\o{\omega}

\def\s{\sigma}

\def\D{\Delta}

\def\L{\Lambda}

\def\S{\Sigma}

\def\Nside{N_{\rm side}}

\def\red{\textcolor{red}}

\def\ha{\frac{1}{2}}

\def\red{\textcolor{red}}

\def\orange{\textcolor{orange}}
\def\cyan{\textcolor{cyan}}

\begin{document}
\title{Local patch analysis of ACT DR6 convergence map using morphological statistics}


\author{Masroor Bashir$^{1,2}$} 
\email{masroor.bashir@iiap.res.in}

\author{Nidharssan S$^{1}$} 
\email{nidharssan@gmail.com}

\author{Pravabati Chingangbam$^{1,2}$} 
\email{prava@iiap.res.in}

\author{Fazlu Rahman$^{3,4}$} 
\email{fazlu@tamu.edu}

\author{Priya Goyal$^{5}$} 
\email{priyagoyal@kias.re.kr}

\author{Stephen Appleby$^{6,7}$} 
\email{stephen.appleby@apctp.org}

\author{Changbom Park$^{5}$} 
\email{cbp@kias.re.kr}

\affiliation{$^1$ Indian Institute of Astrophysics, Koramangala II Block,      
  Bengaluru  560 034, India\\
  $^2$ Department of Physics, Pondicherry University, R.V. Nagar, Kalapet, Puducherry  605 014, India\\
  $^3$ Mitchell Institute for Fundamental Physics and Astronomy, Texas A\&M University, College Station, TX 77843, USA \\
  $^4$ Department of Physics and Astronomy, Texas A\&M University, College Station, TX 77843, USA \\
  $^5$ Korea Institute for Advanced Study (KIAS), 85 Hoegiro, Dongdaemun-gu, Seoul 02455, Republic of Korea \\
  $^6$ Asia Pacific Center for Theoretical Physics, Pohang, 37673, Republic of Korea \\
  $^7$ Department of Physics, POSTECH, Pohang, 37673, Republic of Korea}
\begin{abstract}
We perform a comprehensive hierarchical, multi-scale morphological analysis to search for anomalous behavior in the large-scale matter distribution, using the convergence map from the Atacama Cosmology Telescope Data Release 6 (ACT DR6). Our analysis employs a suite of morphological statistics, including Minkowski functionals, contour Minkowski tensors, and Betti numbers, and measure their deviations from ensemble expectations and median values obtained from isotropic $\Lambda$CDM simulations provided by ACT. 
To assess the statistical significance of these deviations, we develop a general methodology based on the persistence of anomalies across threshold ranges and spatial resolutions, while accounting for correlations among the statistics. 
Analyzing both the full dataset and hemispherical subregions, we find overall consistency with isotropic $\Lambda$CDM expectations. Recognizing that potential deviations in smaller sky regions may be diluted when averaged over larger areas, we further examine localized sky patches. This patch-wise analysis identifies several regions exhibiting statistically significant deviations, which we designate as anomalous. 
In particular, we find that positive density fluctuations near the CMB Cold Spot are anomalous, exceeding 99\% confidence level. This region also overlaps with a previously reported anomalous southern spot in the Planck CMB temperature data. Analysis of the corresponding null map indicates potential contamination from residual foregrounds or instrumental systematics. We also compare the anomalous patches identified in ACT data with those from a previous analysis of the Planck convergence map. No overlapping anomalous regions are found between the two datasets, suggesting that the anomalies in the Planck data are likely attributable to noise.
Further investigation of these atypical patches, considering the role of residual foregrounds, instrumental systematics, reconstruction noise, and independent large-scale structure surveys, is warranted to determine their physical origin.
\end{abstract}

\maketitle

\section{Introduction}

The cosmic microwave background (CMB) photons are weakly lensed by the gravitational potential field  
they encounter along their paths~\cite{Blanchard:1987, Seljak:1995ve, Metcalf:1997ih}.  This results in small but observable distortions in the apparent positions and shapes of CMB temperature and polarization anisotropies \cite{Bartelmann:2001}. This phenomenon of CMB lensing was first detected by cross-correlating CMB data from WMAP with NVSS radio galaxy counts \cite{Smith:2007}. Since the first detection, it has been measured with increasing precision by various CMB experiments such as Planck~\cite{Planck_lensing:2014,Planck:2020}, South Pole Telescope~\cite{Story:2015} and Atacama Cosmology Telescope (ACT)~\cite{Engelen:2015, Darwish:2020, Madhavacheril:2024}.  
These experiments provide measurements of the projected two-dimensional distribution of matter (dark matter and baryons), reconstructed from CMB data,   which can be further used to probe the properties of matter distribution and test different cosmological models~\cite{Hirata:2003,Hanson:2010,Lewis:2006}. 

This data is usually presented in the form of the {\em convergence map}, which is related to the gravitational potential via the Poisson equation. 
This data complements direct observations of the matter distribution via biased tracers such as galaxy distributions. 

Our current theoretical understanding of the universe is based on the assumption of statistical homogeneity and isotropy on large scales, known as the cosmological principle (CP). These assumptions have been supported by observational data and subjected to various tests using diverse cosmic tracers such as X-ray background and counts of radio sources \cite{Alonso:2015ef, Colin:2017gh, Beng:2018bp, Bengaly:2018ba, Rameez:2018ez}, 
clustering properties of galaxies in large scale structure surveys such as Wiggle-Z \cite{Scrimgeour:2012} and SDSS \cite{Ntelis:2017nrj, Sarkar:2018smv,2024MNRAS.530.2866G}. The CMB provides the most conclusive evidence of statistical isotropy (SI) of the Universe as demonstrated through conventional statistical methods such as the power spectrum and N-point functions~\cite{Planckcol:2016bca, Schwarz:2016abs}, as well as more advanced harmonic-space-based techniques like BiPolar Spherical Harmonics (BiPoSH)~\cite{Hajian:2005,Basak:2006,Dipanshu:2023}. A comprehensive review of the status of different cosmological tests can be found in~\cite{Aluri:2023}.  However, hints of deviation from SI have been reported in the literature~\cite{Aluri:2023,Secrest:2021,Secrest:2022,Bashir:2026,Bohme:2018,Khan:2022lpx}. Given these hints of anisotropy, and that different datasets contains different systematic errors, noise  and other uncertainties, it is crucial to test cosmological model assumptions using a diverse range of datasets and statistical techniques.

Real space based morphological statistical tools, namely scalar Minkowski functionals (MFs) \cite{Minkowski:1903, Gott:1990, Mecke:1994} and Minkowski tensors (MTs) \cite{Alesker:1999, Beisbart:2001,Schroder2D:2009,Schr_der_Turk:2013}, provide a methodology to test the $\Lambda$CDM model, and the underlying assumption of statistical isotropy.  In particular, for two-dimensional datasets, the $\alpha$-statistic obtained from the rank-2 contour Minkowski Tensor (CMT) gives a measure of the alignment and alignment of structures~\cite{Chingangbam:2017uqv,Prava:2021}, which can be exploited to search for anomalous orientations in the CMB data~\cite{Ganesan:2017, Joby:2018, Joby:2020,Goyal:2021}. MFs and MTs have also been used to probe other cosmological and astrophysical fields~\cite{Schmalzing:1997uc,Hikage:2012,Munshi:2013,Planck:2015_XVII,Marques:2018ctl, Kapahtia:2018,Kapahtia:2019,Goyal:2019vkq,Appleby:2020,Appleby:2021,Kapahtia:2021, Rahman:2021,Rahman:2024,Martire:2023,Chingangbam:2024,Sabyr:2024bao,PanExGroup:2025}.  Moreover, higher rank Minkowski tensors have also been utilized to search for anomalies in CMB temperature  field~\cite{Collischon:2024}.  
Additionally, Betti numbers are topological quantities whose alternating sum gives the Euler characteristic. Betti numbers for Gaussian random fields were studied in~\cite{Park:2013}, and has been used as a statistical tool in a variety of cosmological studies~\cite{Chingangbam:2012,2017MNRAS.465.4281P,Shukurov:2018,Pranav:2019,Giri:2021,Tymchyshyn:2023czh}. 
They were also recently applied to investigate anomalies in the CMB temperature field~\cite{Pranav:2023}.

In \cite{Goyal:2021}, statistical isotropy (SI)  of the convergence map provided by Planck~\cite{Planck:2020} was tested using the $\alpha$ statistic, and localized anisotropic (anomalous) sky regions were identified. Since the signal-to-noise ratio (SNR) of the Planck convergence data is only order one for low multipoles, and is dominated by noise at higher multipoles, the authors concluded that the anomalous behavior is likely to originate from improper noise estimation. The convergence map provided by the recent ACT  Data Release 6 (DR6)~\cite{Madhavacheril:2024} covers approximately 23\% of the sky. 
This convergence data is signal dominated for multipoles $L < 150$, compared to $L < 90$ for Planck, and has higher  SNR owing to higher precision and better noise mitigation techniques.  Hence, the ACT data serves as a new, improved testbed for the standard isotropic $\L$CDM cosmology. In this paper, we extend the methodology of  \cite{Goyal:2021}  to include MFs and Betti numbers in addition to $\a$, and analyze the convergence data from the ACT DR6, to search for anomalous deviations of the data from standard isotropic $\L$CDM cosmology. 
We restrict our analysis to an appropriate range of scales since the data for $L > 1300$ may not be reliable due to the presence of secondary anisotropies which bias the lensing reconstruction, as pointed out in \cite{MacCrann:2024}.  
Our analysis involves studying the convergence map at multiple scales, by extracting the MFs, CMT and Betti numbers from the global (masked) map, the north and south galactic sky separately, and from individual patches of various sizes. 
This work constitutes a search for regions of the sky that exhibit anomalous statistical properties, through comparison of data with statistically isotropic $\L$CDM simulations. Such a test is motivated by previously discovered curiosities; the anomalous hemispherical power asymmetry \cite{Eriksen:2003db,Eriksen:2007pc,Planckcol:2016bca,Schwarz:2015cma,Akrami:2014eta,Sanyal:2024iyv} is an example of such a feature, and on smaller scales, the Cold Spot is also an anomalous local region \cite{Vielva:2004}.

This paper is organized as follows. Section~\ref{sec:sec2} reviews the basic physics of CMB lensing, and the ACT data that we analyze is presented. Section~\ref{sec:sec3} gives a brief description of the morphological statistics we use. Section~\ref{sec:sec4} discusses the issue of noise mitigation of the data. Our results for the morphological statistics are presented in Section~\ref{sec:sec5}. Supplementary materials are included in appendices~\ref{sec:a1} to 
\ref{sec:a5}. 
We end with a summary and discussion of our results in Section~\ref{sec:sec7}.

\section{Physics of CMB lensing and ACT data set}
\label{sec:sec2}

The lensed CMB temperature or polarization field is related to the unlensed field through a remapping,
\begin{align} 
   T_{\rm lensed}(\vec{n}) &= T_{\rm unlensed}(\vec{n} + \vec{\alpha}),\\
   (Q + iU)_{\rm lensed}(\vec{n}) &= (Q + iU)_{\rm unlensed}(\vec{n} + \vec{\alpha}),
   \label{eqn:lens_eqn}
\end{align} 
where $\vec{\alpha}$ is the deflection angle. The deflection angle for small perturbations, under the Born approximation can be expressed as the gradient of the lensing potential, $\Psi$: $\vec{\alpha}\,(\hat{n})= \vec{\nabla} \Psi \, (\hat{n})$. 
The lensing potential ($\Psi$) represents the integrated gravitational potential of all matter distributed along the line of sight to the last scattering surface (LSS)~\cite{Smith:2007rg}. 
Another key observable in CMB lensing studies is the lensing convergence, $\kappa \, (\hat{n})$, which in accordance with Poisson equation is given by,
\begin{eqnarray} 
\kappa \, (\hat{n}) = -\frac{1}{2} \Vec{\nabla}_{\hat{n}} ^{2} \Psi \, (\hat{n})  = -\frac{1}{2} \Vec{\nabla}_{\hat{n}} \cdot \vec{\alpha} \, (\hat{n}) ~.
\end{eqnarray}
$\kappa$ captures the local magnification or de-magnification of the observed CMB fields caused by the matter distribution. Hence, convergence directly probes the total mass density of the universe integrated along the line-of-sight all the way to the redshift of recombination, $z_* \approx$ 1100 , with a peak contribution from around $z\sim 2$~\cite{Madhavacheril:2024,Qu:2024, Darwish:2020}. Alternatively, $\kappa$ can also be expressed as an integral over the matter distribution $\delta_m$	
(the matter over-density) along the line of sight through: 

\be
\kappa(\hat{n}) = \int_0^{\infty} dz \, W_\kappa(z) \, \delta_m(\chi(z) ,\hat{n}).
\ee
In the case of a flat universe, the lensing kernel \( W_\kappa \) simplifies to:
\be
W_{\kappa}(z) = \frac{3}{2} \Omega_m H_0^2 \frac{(1 + z)}{H(z)} \frac{\chi(z)}{c} \left( \frac{\chi(z_*) - \chi(z)}{\chi(z_*)} \right),
\ee
where \( \chi(z) \) is the comoving distance to redshift \( z \).

$\kappa$ or $\psi$ can be reconstructed from the observed lensed CMB fields using quadratic estimators~\citep{Zaldarriaga:1998te,Okamoto:2002,Okamoto:2003kl}. There are other methods developed for extracting the lensing potential field from CMB like gradient inversion method~\citep{2019PhRvD.100b3547H}, or, likelihood based methods~\citep{PhysRevD.67.043001,Hirata:2003,Carron:2017}. However, the quadratic estimators are known to be sub-optimal especially for polarization and  hence, remain as the most commonly employed reconstruction methodology for various recent and upcoming CMB experiments like Planck~\citep{Aghanim:2018oex}, ACT~\citep{Qu:2024}, Simons Observatory~\cite{SO_2019JCAP}, CMB S4~\cite{CMB-S4:2016ple} etc.

\subsection{ACT DR6  convergence data}
\label{sec:sec2b}

The ACT has mapped CMB lensing over $9400\,{\rm deg}^2$, achieving a state-of-the-art 2.3\% precision in the lensing power spectrum amplitude~\citep{Qu:2024,Madhavacheril:2024,MacCrann_2024}.
This map has been recently released as a part of ACT DR6 and is based on the CMB measurements made between $2017$ and $2021$ at $90$ and $150$ GHz. 

 The ACT lensing map reconstruction methodology employs a cross correlation based estimator introduced in~\cite{Madhavacheril:2021}. This estimator utilizes multiple time-interleaved splits of the CMB maps rather than using a single map. This ensures that the instrumental noise associated with each individual map is independent. As a result, the CMB lensing map derived from this method is not susceptible to the modeling of instrumental noise. The lensing map is reconstructed with CMB scales from $600 < \ell < 3000$ and is signal dominated on scales $L <150$. The large CMB scales are excluded due to significant contribution from atmospheric noise, Galactic foregrounds and a correction greater than $10\%$  
 for the large scale transfer functions, whereas the small scales are excluded to minimize contamination from astrophysical foregrounds like the thermal Sunyaev-Zeldovich (tSZ) effect, the cosmic infrared background (CIB), and radio sources~\citep{MacCrann_2024}. To further mitigate the effects of extragalactic foregrounds, the lensing reconstruction uses a profile-hardened lensing estimator~\citep{Sailer:2020}. This involves constructing a quadratic estimator that is immune to the contribution to the CMB mode coupling arising from objects with radial profiles similar to those expected from tSZ clusters. 

We analyze the lensing convergence ($\kappa$) map derived from ACT temperature and polarization data, employing a profile-hardening estimator for better foreground mitigation. This map is referred to as a "baseline map". The ACT team provides spherical harmonic modes, $\kappa_{LM}$ of the $\kappa$ map in the \texttt{healpy} indexing scheme. They correspond to modes of a map described in equatorial coordinates, as opposed to the Galactic coordinates used in Planck products. The ACT data is significantly improved over Planck data,  particularly in angular resolution ($0.5'$) and in reducing reconstruction noise power by at least a factor of two. Although the Planck maps cover more than twice the area, the ACT maps reconstruct small scales with greater precision, making them especially valuable for cross-correlation studies with galaxy groups~\citep{Farren:2024}.


{\em Simulated convergence maps}: The ACT DR6 release also includes 400  realistic simulations essential for validating lensing data analysis, testing systematic errors, and refining cosmological methods using CMB lensing maps. The process begins with generating input Gaussian lensing convergence fields based on a fiducial lensing convergence power spectrum. These fields represent the expected gravitational lensing effects on the CMB due to large-scale structures. Lensing is then applied to unlensed Gaussian CMB temperature and polarization maps, producing lensed CMB maps that mimic real sky observations. These lensed maps are further processed through the ACT simulation pipeline, incorporating instrumental effects and noise typical of the ACT survey. Finally, the mock datasets undergo CMB lensing reconstruction, simulating the actual observational data analysis. 
Both the observed and simulated convergence maps are provided as FITS files containing the spherical harmonic modes $\kappa_{LM}$ of each map with $L_{max}=4000$, in equatorial coordinates.

{\em Baseline mask:} In our analysis, we use the "baseline" ACT DR6 mask in \texttt{HEALPix} format. This mask covers  $\sim 23 \%$ of the sky, assigning a value of $1$ to the selected regions and smoothly transitions to $0$ near the edges. It is specifically designed to avoid the dusty regions, ensuring that our analysis focuses on cleaner regions of the sky with minimal contamination from galactic dust. The mask is consistently applied to both observed and simulated data, with the appropriate $\Nside$ resolution.

{\em Null maps}: ACT DR6 includes two null maps constructed by differencing the CMB maps at 90 GHz and 150 GHz. The first null map is based solely on CMB temperature data, while the second incorporates both temperature and polarization measurements.
 For our analysis, we use the latter. Null maps are expected to be free of cosmological signals due to cancellation in the subtraction. What is then left behind are the following components. First, foregrounds signals like the cosmic infrared background  and the thermal Sunyaev–Zel’dovich effect which do not fully cancel out because they have distinct spectral characteristics at 90 GHz and 150 GHz. Second, since the beam profiles at the two frequencies differ, the null maps may still contain residual beam-related systematics. Third, the instrument noise at the two frequencies can vary and may not cancel out completely.  In addition to the null maps obtained from observed data, the DR6 dataset also includes 400 simulated null maps, which are useful  for validating the observational null maps. By comparing observed and simulated null maps, one can check whether any features that persist in the observed convergence maps are likely due to residual foregrounds, instrument related systematics and noise, or just random statistical fluctuations.

\section{Methodology: Minkowski Functionals, Minkowski Tensors, and Betti numbers}
\label{sec:sec3}

The geometry and topology of a smooth random field is usually studied by constructing excursion sets~\cite{adler2010} which comprise of spatial points where the field has values above the chosen field level. 
Let $Q^{\nu}$ denote an excursion set indexed by field level $\nu\sigma_0$, $\sigma_0$ being the standard deviation of the field.  
Then for a field on flat 2D space, Minkowski tensors are defined as the following set of functionals:
\begin{eqnarray}
V_0^{i,j} &\equiv& \frac{1}{A}\int_{Q^{\nu}} \vec r^i\otimes \hat n^j\da, \\
V_1^{i,j} &\equiv& \frac{1}{4A}\int_{\partial Q^{\nu}} \vec r^i\otimes \hat n^j\ds, \\
V_2^{i,j} &\equiv& \frac{1}{2 \pi A}\int_{\partial Q^{\nu}} \vec r^i\otimes \hat n^j\ \k\,\ds,
\label{eq:mts}
\end{eqnarray}
where $\da$ is the area element, $\ds$ is the element of the arc length along the excursion boundary $\partial Q^{\nu}$, $\otimes$ is the symmetric tensor product, $\vec r^i$ is the tensor product of $i$ copies of the position vector, $\hat n^j$ is the tensor product of $j$ copies of the unit normal vector to the excursion set boundary, and $\k$ is the geodesic curvature\footnote{Note that in this section, we use $\kappa$ to denote geodesic curvature. In other sections, $\kappa$ refers to the lensing convergence map.}. $A$ in the prefactors denote the total area of the space over which the field is defined. This prefactor makes the MTs in eq. (\ref{eq:mts}) to be defined as densities. 
The rank zero MTs corresponding to the case $i,j=0$ give the three scalar Minkowski functionals~\cite{Mecke:1994,Gott:1990,Schmalzing:1997,Schmalzing:1997uc},
\begin{eqnarray}
V_0(\nu) &=& \frac{1}{A}\int_{Q_{\nu}} \da,\\
V_1(\nu) &=& \frac{1}{4A} \int_{\partial Q_{\nu}} \ds, \\
V_2(\nu) &=& \frac{1}{2\pi A} \int_{\partial Q_{\nu}} \k\,\ds.
\end{eqnarray}
For rank $2$, with $i+j=2$, there are three independent translation invariant MTs whose traces give the three MFs. Out of these, only 
\be V_1^{0,2} =\frac{1}{4A} \int_{\partial Q_{\nu}} \hat n\otimes \hat n \,\ds,\ee
contains additional morphological information beyond the scalars.  
From the eigenvalues, $\L_1, \L_2$, of $V_1^{0,2}$, we can define a measure of anisotropy and relative orientation~\cite{Schroder2D:2009,Chingangbam:2017uqv, Appleby:2017uvb, Prava:2021} as
\begin{equation} \label{eq:alpha}
    \alpha=\frac{\Lambda_{1}}{\Lambda_{2}}.
\end{equation}
By definition we have $0<\a\le 1$. If the iso-field contours (excursion set boundaries) are randomly oriented with no preferred direction, and we take the area of the manifold on which the field is defined to infinity, then we expect $\alpha \to 1$. Any preferential alignment will give $\alpha<1$. Departures from $\alpha = 1$ are also expected due to statistical noise arising from the finite extent of cosmological data. The value of $\alpha$ therefore provides an intertwined measure of the degree of departure from isotropy and statistical noise. 

{\em Minkowski tensors for fields on the unit sphere}: Since in this paper we focus on the convergence field which is  given on the unit sphere, the statistics we use must be defined on the sphere. The definitions of translation invariant MTs of rank zero and two can be naturally extended to fields on curved spaces such as the unit sphere~\cite{Chingangbam:2017uqv}, which is relevant to our study here. However, for data defined on the two-sphere, and more generally on a curved manifold, the definition of Minkowski tensors as spatial averages is ambiguous and the concept of alignment between structures is subtle \cite{Appleby:2022itn}. For fields that occupy a significant fraction of the sphere, performing a spatial average of the vector product $\hat{n} \otimes \hat{n}$ will inevitably lead to a rotation of the vectors when transporting them. We can expect such a rotation to artificially isotropize the measurements, making $\alpha$ problematic to interpret when searching for global isotropy. In this work, we will be predominantly concerned with statistics extracted from small patches on the sky, where the curvature associated with the manifold is less significant. Effectively, for the purpose of extracting information from the tensors, we will use the flat sky approximation within each patch.

{\em Numerical computation of scalar MFs and CMT}: We adopt the method outlined in~\cite{Chingangbam:2017uqv} which builds on~\cite{Schmalzing:1997uc}.  $V_{0}$ is  obtained by using Heaviside step function to define the pixels above the field level belonging to each excursion set.  For $V_1, V_2$ and $V_1^{0,2}$, we express $\hat n$ and $\k$ in terms of the first and second derivatives of the field. Then the line integral is converted to an area integral by incorporating a Jacobian factor and $\delta$-function. Numerical approximation of $\delta-$function leads to small numerical error which can be quantified and corrected as described in~\cite{Lim:2011kd,Prava:2021}.  

{\em Numerical computation of  Betti numbers}: We identify the distinct connected regions and holes in the excursion set. This involves labeling each pixel in the field according to its membership to a specific subset. All pixels are checked sequentially; if the field value at the $i^{\rm th}$ pixel 
lies above the threshold $\nu$ then a new connected component is defined. All pixels in the neighbourhood of the $i^{\rm th}$ pixel where the field values 
similarly lie above the threshold, are included in this connected component; and neighbours of neighbours are iteratively checked until no more pixels are found satisfying the two criteria (adjacent to a pixel in the newly defined connected component, field value above the threshold $\nu$). The sequential check through the pixels is then resumed until the next pixel is found that is above $\nu$ and does not already belong to a previously defined connected component. The Betti number $b_{c}$ is the total number of excursion set regions found using this method. The algorithm is repeated for all holes (regions that satisfy $\delta < \nu$) and $b_{h}$ is the number of distinct regions found below the threshold.   
Excursion sets and holes that touch the boundary of the field can introduce artifacts that distort the true count. To address this, we apply a correction procedure that removes these boundary-touching regions from consideration. This ensures that only those connected regions and holes entirely contained within the field contribute to the Betti numbers. The price we pay is that we expect a systematic decrease in the Betti numbers for regions of the sky that are more heavily masked. However, because we apply the same mask and methodology to the mock data to which we compare, we do not expect any anomalous results due to our approach. 

Lastly, note that MFs and MTs are usually defined as surface densities, i.e. per unit area. We adopt this convention as well. Before computation of morphological statistics for each field, we first subtract the mean and divide by the standard deviation. The morphological statistics are then computed in the threshold range $-4 < \nu < 4$ with spacing $\Delta\nu=0.2$.

\subsection{Quantifying the uncertainties of morphological statistics}
\label{sec:sec3c}

 In the next section, morphological statistics computed from ACT data and simulations will be compared. To quantify the significance of any disagreement between simulated and actual data, we need to build the probability distribution function of the summary statistics, including the covariance between different statistics computed at various threshold values. In cosmological applications, it is usual to assume that the Minkowski functionals have Gaussian distribution.  This is reasonable when the number of structures is large and the field is dominated by small-scale modes\footnote{The precise nature of uncertainties is not known analytically. The presence of large scale modes modifies the distribution function of summary statistics, see for example \protect\cite{Bernardo:2024uiq} (Appendix A).}. Likewise, Betti numbers are also assumed to be Gaussian. In \cite{Chingangbam:2025}, it has been shown that Betti numbers and the Euler characteristic (equivalent to $V_2$) can be modelled as sums/differences of Binomial variables, and are Gaussian to a good approximation for threshold ranges upto  a few standard deviations. For these reasons, in this study we also work with this assumption.  The remaining statistic $\a$, obtained from the CMT, can be modeled as following the Beta distribution~\cite{Goyal:2021}.

Let $X_{\mu, i}$ denote a measurement of one of the statistics  ($V_{0}, ~V_{1},~ V_{2}, b_c$, $b_h$, $\a$) at a particular threshold, where the subscript $\mu$ 
identifies the statistic, and $i$ denotes the $\nu_{i}$ threshold bin\footnote{We do not use the repeated index sum convention in this section.}.  For each $\mu$, let $\bar X_{\mu, i}$ and $\s_{X_{\mu,i}}$ respectively denote the mean and standard deviation of $X_{\mu,i}$ computed from the 400 simulations provided by ACT (see section \ref{sec:sec2b}). We define the difference between the data-measured value $X_{\mu, i}^{\rm obs}$, where the superscript `obs' refers to the baseline map, and $\bar X_{\mu, i}$ as, 
\be
\D {X}_{\mu,i} (\nu_{i}) = X_{\mu, i}^{\rm obs} (\nu_{i}) - \bar{X}_{\mu, i} (\nu_{i}), 
\label{eq:Xidev} \ee 
and finally define the $\chi_\mu^2$ variable as, 
\be
\chi_{\mu}^2 =  \sum_{j,k} \D X_{\mu,j} \left(\S_\mu^{-1}\right)_{jk} \D X_{\mu, k},
\ee
where the indices $j,k$ run over the number of threshold values at which $X_{\mu,i}$ is computed and $\S_\mu$ is the covariance matrix for each $\mu$ (see Appendix \ref{sec:a1}). For our global and hemispherical analysis of the data, we measure $\chi_{\mu}^{2}$ individually for each statistic.

In what follows, we also search for local anomalies in the data by measuring the quantities ($V_{0}, ~V_{1},~ V_{2}, b_c$, $b_h$, $\a$) in patches on the sky. When doing so, for clarity, we combine the morphological statistics into one large data vector rather than assess each $\chi_{\mu}^{2}$ individually. Because the summary statistics are correlated, we cannot simply sum the $\chi^{2}_{\mu}$ values within each patch. Instead we define a large array $X_I$ that contains the statistics ($V_{0}, ~V_{1},~ V_{2}, b_c$, $b_h$) at all density thresholds, so $I$ subscripts run over $5 \times N_{\nu}$, where $N_{\nu}$ is the total number of $\nu_{i}$ threshold bins. We can  define the combined $\chi^2$ variable
\be
\chi^2 =  \sum_{J,K} \D X_J \left(\S^{-1}\right)_{JK} \D X_K,
\label{eq:Xchi}
\ee
where the indices $J,K$ run over all the statistics and all the threshold values at which they are measured, and $\S$ is the corresponding covariance matrix. Note that we do not include $\alpha$ in the array $X_{I}$; it is treated separately as discussed below.

For those $X_{\mu, i}$ quantities that are approximately Gaussian, we can treat $\chi_{\mu}^2$ as a chi-squared distributed variable and use it to evaluate the statistical significance of deviations of $\D {X}_{\mu, i}$ from zero. However, this approach requires an accurate reconstruction of $\Sigma_{\mu}$, which is not feasible given our dataset of $400$ simulations only. For this modest number, the inverse $\Sigma_{\mu}^{-1}$ will not converge to an unbiased estimate of the `true' underlying inverse covariance matrix. To resolve this problem, we model our dataset using a modified multivariate $t$-distribution rather than  multivariate Gaussian, thereby accounting for the uncertainty arising from the small sample size. This accurately represents the multivariate structure in cases where the covariance matrix is estimated from the same dataset \cite{Sellentin:2017}. When the PDF is multivariate $t$-distribution, the $\chi^2$ variable defined in eq. (\ref{eq:Xchi}) follows a scaled $F$-distribution. We have numerically verified this, and shown that this tends to a standard chi-squared distribution as the sample size increases. 
More details about the method are given in appendix \ref{sec:a11}. 

For $\alpha$, we quantify deviations using the $\mathcal{M}^2$ method (see  Appendix \ref{sec:a12}), which is a modified version of the aforementioned chi-square test to be applicable to Beta-distributed variables. We fit a Beta distribution at each threshold and generate samples to numerically estimate the distribution of the $\chi^2$ variable described in eq. (\ref{eq:Xchi}), which is in turn used to calculate the $p$-value for the observed data. Since the observed correlations across thresholds are small (see figure \ref{fig:cov}), this method neglects them. Further details about this method can be found in Appendix \ref{sec:a12}. 

\subsection{Condition number}  
\label{sec:s3B}

In addition to the issue of non-convergence of the covariance matrix due to the limited number of simulations, as discussed in the previous subsection, it is also essential to ensure that the inversion of the covariance matrix is numerically stable. To assess this, we employ the `condition number' as a diagnostic measure of numerical stability, as described below.

For a generic non-singular square matrix, say $A$, let $s_1, s_2$ denote its largest and smallest singular values\footnote{Singular values are non-negative square roots of the eigenvalues of $A^TA$.}, respectively. Then the condition number, $C$, is defined to be~\cite{belsley:1980},
\be
C=s_1/s_2.
\ee
$C$ provides a measure of how stable the inversion of $A$ is against small changes of its elements. $C=\infty$ when $A$ is singular, in which case $A$ is said to be ill-conditioned.  If $C$ is small and close to one, $A$ is said to be well-conditioned. For a given $A$, while there is no generally agreed upon cut-off value, it is desirable that $C$ is not too large. 

In the subsequent analysis, we compute the condition number $C$ for various covariance matrices corresponding to different statistics, as detailed in Section~\ref{sec:sec5}. While we do not impose a strict cutoff on $C$, a given statistic is used for physical inference only if its associated covariance matrix yields condition numbers well below $10^3$.

\subsection{Note on multiscale analysis}
\label{sec:s3C}

Probing multiscale information can be carried out in two ways. One approach is to fix $N_{\rm side}$ and vary the smoothing scale, with each scale obtained by a single convolution with a Gaussian kernel. Alternatively, one can vary $N_{\rm side}$ by downgrading  and apply the corresponding smoothing. In this case, the map is first convolved with a top-hat kernel associated with the change in $N_{\rm side}$, followed by a convolution with a Gaussian kernel of the desired smoothing scale. 

Although these procedures involve different averaging schemes, they are expected to yield similar, though not identical, results. We have verified that the two coarse-graining procedures give consistent morphological statistics. In this work, we adopt the second approach because it is substantially more computationally efficient. We characterize the probed scales by the full width at half maximum (FWHM) of the Gaussian smoothing kernel.

\section{Morphological statistics for ACT baseline map - noise mitigation}
\label{sec:sec4}

The ACT DR6 convergence map is increasingly more noise dominated towards smaller scales. See figure 5 of \cite{Madhavacheril:2024} for the noise power spectrum, which also includes the Planck noise power spectrum for comparison.  
To mitigate this effect of reconstruction noise, we apply Wiener filter    
to the spherical harmonic coefficients $\kappa_{LM}$, as given by the following expression,
\be
\kappa_{LM}^{WF} = \frac{C_L^{\kappa \kappa}}{C_L^{\kappa \kappa} + N_L^{\kappa}} \,\kappa_{LM}.
\label{eq:WF}
\ee
Here, $C_L^{\kappa \kappa}$ is the convergence power spectrum in the fiducial cosmological model based on the measurement of the primary CMB anisotropies by the Planck collaboration \cite{Aghanim:2020}, and $N_{L}^{\kappa}$ is the reconstruction noise per mode power spectrum given upto $L_{\rm max}=2101$. Let us denote $w_L= C_L^{\k \k}/\left(C_L^{\k \k} + N_L^\k\right) $.  From  figure 5 of \cite{Madhavacheril:2024}, we see that $C_L^{\k \k} \sim N_L^\k$ around $L\sim 150$. Hence, for $L> 150$, $w_L \propto (N_L^\k)^{-1}$, so $\kappa_{LM}^{WF}$ is suppressed with respect to $\kappa_{LM}$. On the other hand, for $L<150$ we have $w_L\to 1$, which implies  $\kappa_{LM}^{WF} \sim \kappa_{LM}$.

\begin{figure*}[htp]
    \centering 
    With Wiener filter \hskip 4.5cm Without Wiener filter \\
   \includegraphics[width=0.8\textwidth]{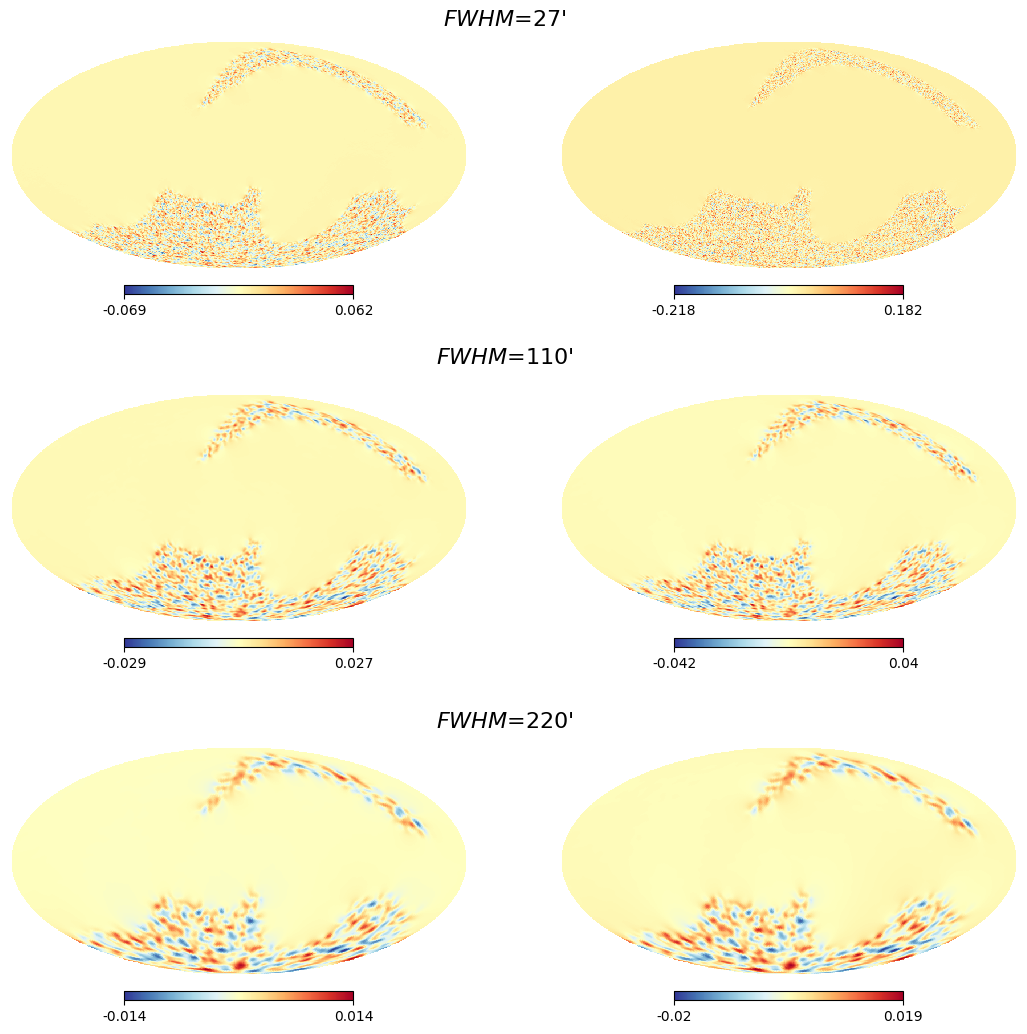}
   
    \vskip .3cm
    \includegraphics[width=0.8\textwidth]{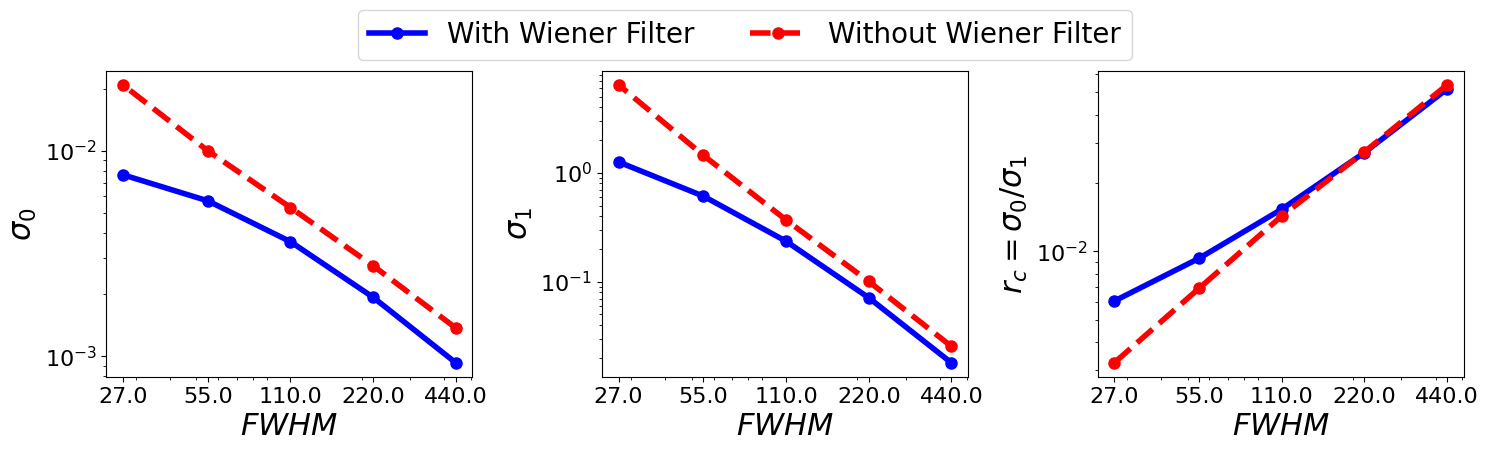}
    \caption{{\em Rows 1 to 3}: ACT baseline convergence map with (left) and without (right) Wiener filtering for different FWHM 
    values, shown in Galactic coordinates. {\em Bottom row}: $\sigma_0$, $\sigma_1$ and $r_c$ for maps with and without WF, as functions of FWHM 
    The value of $r_c$ becomes comparable for FWHM$\gtrsim 110'$.}
    \label{fig:act_baseline}
\end{figure*}
We generate two sets of $\kappa$ maps, with and without Wiener filtering,  from $\kappa_{LM}$ and $\kappa_{LM}^{WF}$ at different $\Nside$ values with corresponding $L_{\rm max}=3N_{\rm side}-1$. 
To ensure smooth field derivatives, we further apply Gaussian smoothing with appropriate FWHM values  
of $27',\, 55'$ and $110'$ 
for $\Nside=512, \ 256$ and 128, respectively. This corresponds to FWHM values spanning roughly four pixels for each $\Nside$.   
In what follows, we will quote the scale probed in terms of FWHM.  Figure~\ref{fig:act_baseline} shows the ACT baseline convergence map at different FWHM 
values  in Galactic coordinates. The left column corresponds to Wiener filtered maps while the right column shows the maps without Wiener filtering.  Note that the ranges of the colorbars differ between left and right columns, with the left being smaller due to $w_L$ being less than one. At  FWHM=$27'$ (top row) 
it can be discerned by eye that the unfiltered map is more grainy, indicating higher amplitude of small scale fluctuations, compared to the filtered one. As FWHM increases, 
the filtered and unfiltered maps become comparable in terms of their morphology.

To quantify the visual observations, we compute the standard deviations of each field and its gradient field, denoted by $\s_0$ and $\s_1$, respectively, along with their ratio $r_c=\s_0/\s_1$. These are plotted in the left and middle panels of the last row in figure \ref{fig:act_baseline}. Their ratio $r_c=\s_0/\s_1$, shown in the right panel,  gives a statistical measure of the typical size of spatial fluctuations of the field. As expected, $\s_0$ and $\s_1$ are smaller for Wiener filtered maps compared to the unfiltered ones, for each FWHM. 
We find that $r_c$ is lower for the filtered maps at lower FWHM,  
and becomes comparable with the unfiltered maps roughly for FWHM$\gtrsim 110'$. 
This is due to different scaling of $\s_0$ and $\s_1$ with FWHM 
for the filtered and unfiltered maps.

\begin{figure*}[htp]
   \centering 
    {\bf (A)}\\
    \includegraphics[width=0.7\textwidth]{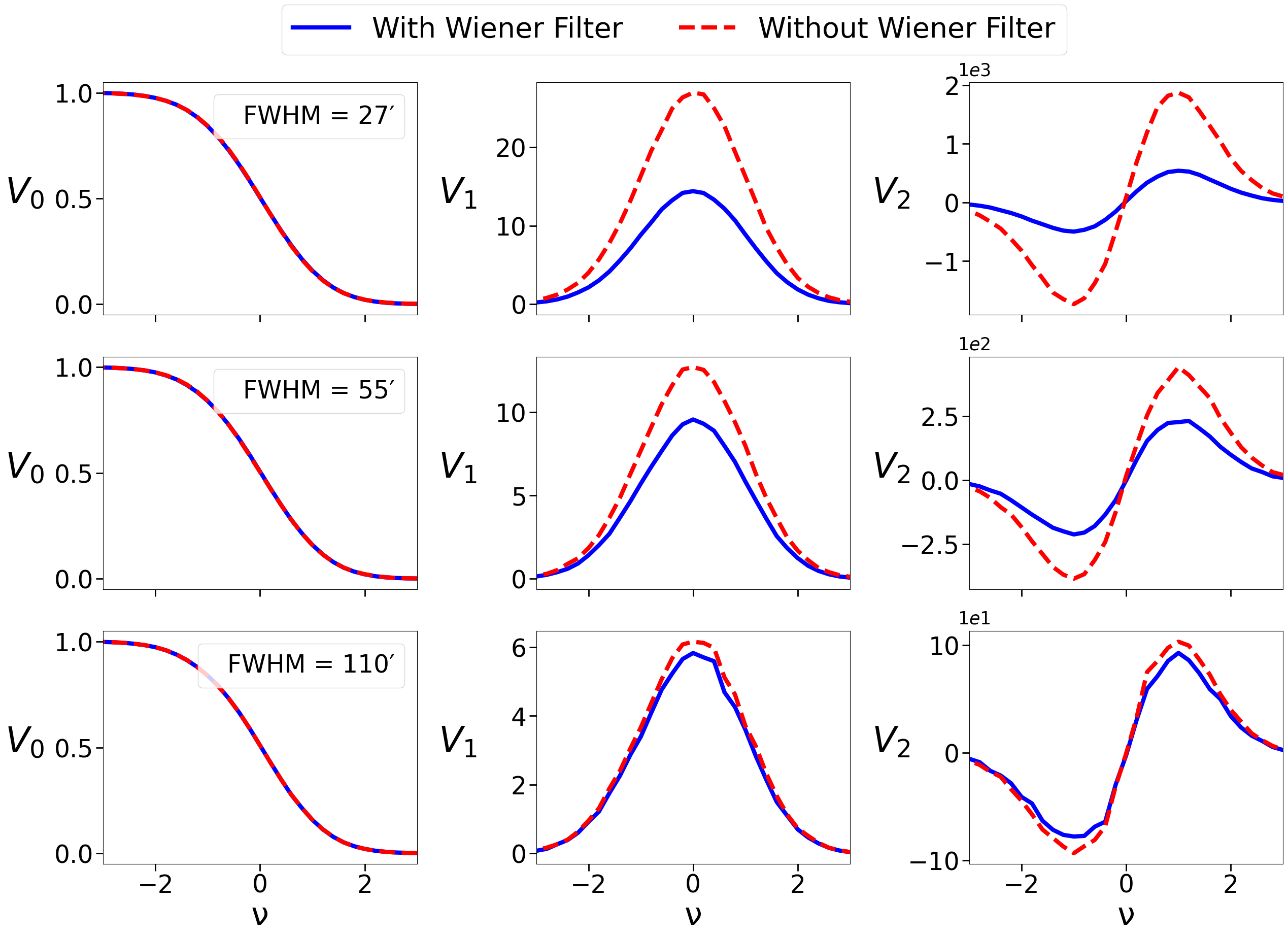}\\
    \vskip .2cm
    {\bf (B)}\\
       \includegraphics[width=0.7\textwidth]{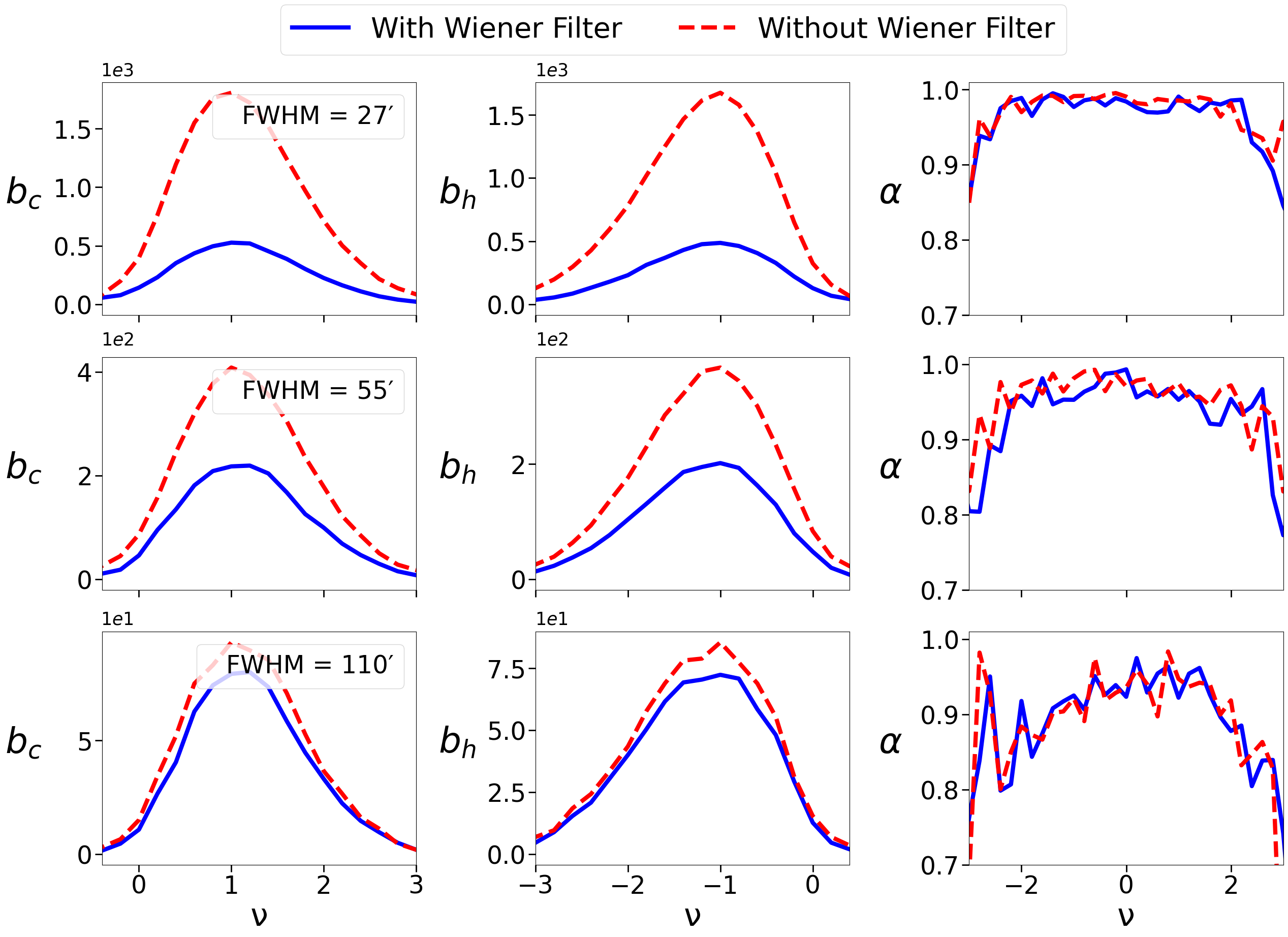}
       \vskip .2cm      
   \caption{{\bf (A)}: $V_{0}$, $V_{1}$, $V_{2}$ versus $\nu$ for different FWHM 
   values of the baseline map, with (blue solid lines) and without (red dashed lines) Wiener filtering. All the panels have same $x$-ranges. {\bf (B)}: $b_h$, $b_c$ and $\alpha$ for the same maps as {\bf (A)}. Note that the $x$-ranges here differ from those in (A). }
    \label{fig:WF}
\end{figure*}


The observed convergence map is the sum of the true convergence map, and contaminants such as residual foregrounds, instrument systematics and noise.  At the angular scales of interest, the true $\kappa$ is expected to be Gaussian to good approximation.  So, if contaminants are Gaussian (which is unlikely), then the observed $\kappa$ would also be Gaussian to good approximation.  
For Gaussian random fields, analytic expressions for MFs are well known \cite{Schmalzing:1997}. $V_0=\ha {\rm erfc}(\nu/\sqrt{2})$ is independent of cosmology, while $V_1\propto r_c^{-1} e^{-{\nu^2}/2}$ and $V_2\propto r_c^{-2}\nu e^{-{\nu^2}/2}$, encode cosmological information through $r_c$.  
Figure~\ref{fig:WF} shows the full set of morphological statistics computed for  different FWHM 
values  for the original ACT baseline convergence map (red dashed lines) and the Wiener filtered map (blue solid lines). 
The behaviour of $V_0, V_1, V_2$ broadly  follows the Gaussian analytic expressions described above. In the following, however we do not attempt to fit the MF curves to their Gaussian predictions, since our aim is not to test for non-Gaussianity but rather to assess anomalies in the observed map 

For $b_h$ and $b_c$, analytic expressions are not known even for the Gaussian random fields. Also, they are more sensitive to noise, and to the presence of boundaries in the data, and hence more difficult to measure compared to $V_{2}$. For most thresholds of interest, we have $b_h \sim V_2$ for $\nu\lesssim -1$ and $b_c \sim V_2$ for $\nu\gtrsim -1$, ignoring the curvature of the sphere. For these reasons, it is more common to study the Minkowski Functional $V_{2}$ than $b_{c}$, $b_{h}$. However, we expect that the Betti numbers contain more information compared to $V_{2}$, because both the shape and amplitude of the Betti curves as a function of threshold are sensitive to the power spectrum of the field \cite{Park:2013}. In contrast, for a Gaussian field, only the amplitude of $V_{2}$ carries such information. 

Next we discuss the statistic $\alpha$. As shown in figure \ref{fig:WF} (B), from the bottom to the top panels,, the number of structures increases by orders of magnitude as the smoothing scale is decreased; this increase is reflected in the values of $b_{c}$ and $b_{h}$. However, the value of $\alpha$ approaches unity very slowly, changing by $\sim 5\%$ (cf. right panels). Also, the number of structures increases when we switch off the Wiener filtering (cf. red dashed lines), but $\a$ is only weakly sensitive to this increase. Clearly, the asymptotic behavior of $\a \to 1$ is slow as the number of structures grows. As $\a$ approaches unity, the underlying probability distribution function from which it is drawn is increasingly non-Gaussian and skewed towards values below unity; this is due to the definition of $\a$ that imposes $\alpha \leq 1$. We also observe less correlation between threshold bins for $\alpha$ compared to all other statistics. This is reflected in the correlation matrices presented below.  

\begin{figure*}[htp]
     \centering 
     Positive thresholds \hskip 2.8cm Negative thresholds \hskip 2.8cm Full threshold range\\
       \includegraphics[width=0.3\textwidth]{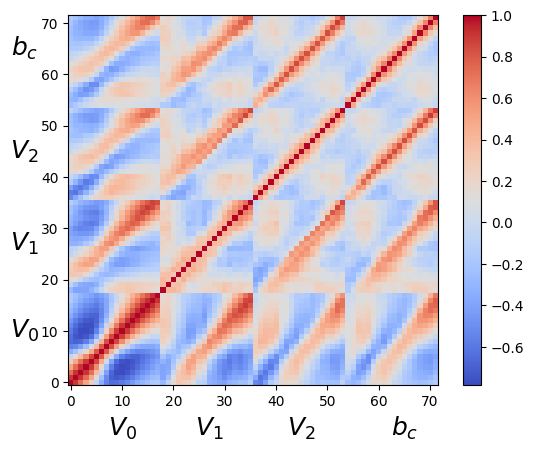} 
         \includegraphics[width=0.3\textwidth]{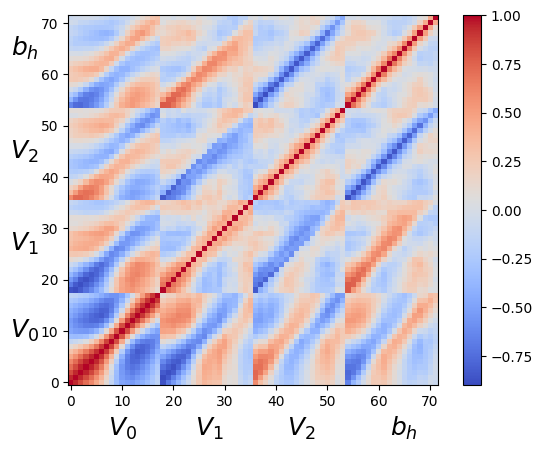}
           \includegraphics[width=0.3\textwidth]{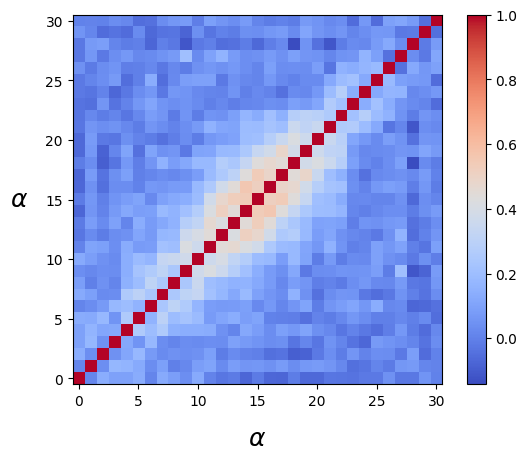}\\
    \caption{Cross-correlation matrices between the statistics across thresholds for FWHM=55', 
    computed from 400 simulated convergence maps. The plot shows drop in cross-correlation from $V_0$ to $V_2$ and Betti numbers, whereas $\alpha$ shows significantly weaker cross-correlations between thresholds with the exception of a small region near $\nu = 0$.}
    \label{fig:cov}
\end{figure*}

{\em Correlation matrix for the statistics}: The correlation matrices (covariance normalized by the respective standard deviations) for the morphological statistics,   computed using the 400 simulations of $\kappa$, are shown in figure~\ref{fig:cov} for FWHM$=55'$. 
The plots change only qualitatively for other FWHM 
values. It is interesting to note a  drop in the degree of (anti-) correlation from $V_0$ to $V_2$, and Betti numbers.  There is a trend; ignoring the area normalization of the statistics, the lower the spatial dimension of the statistic the weaker the cross-correlation between thresholds. For $\alpha$, the colour bar ranges from zero to one, unlike for $S_+$ and $S_-$. We see that $\a$ shows negligible correlations across different threshold values, except in the vicinity of $\nu=0$ where we see positive correlations across a few threshold bins.  

\section{Analysis and results}
\label{sec:sec5}

The results that we present in this section will focus on  Wiener filtered maps. We perform our analysis by hierarchical and multiscale iterations of the sky regions. The analysis is classified as: (1) {\em global} analysis, where we calculate all statistics for the full convergence map, (2) {\em hemispherical} analysis, where we calculate all statistics separately for the northern and southern hemisphere (in Galactic coordinates),  and (3) {\em local or patch} analysis, where we divide the sky into smaller non-overlapping patches and compute the statistics for each patch.

\subsection{Global and hemispherical analysis}
\label{sec:sec5a}

\begin{figure*}[t]
    \centering 
    {\bf (A)}\\
    \includegraphics[width=13cm,height=8cm]{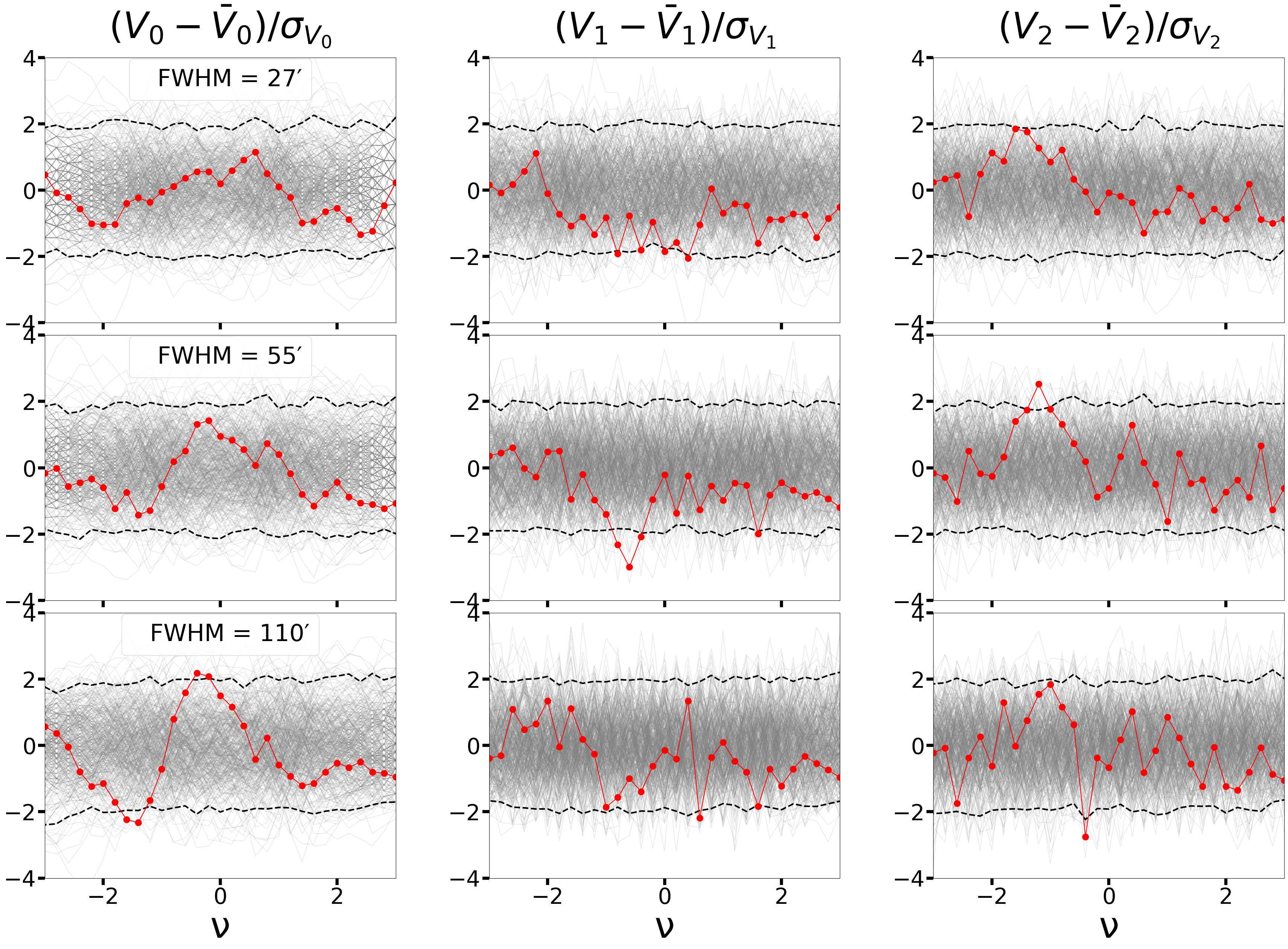}\\
    \vskip .4cm
    {\bf (B)}\\
       \includegraphics[width=13cm,height=8cm]{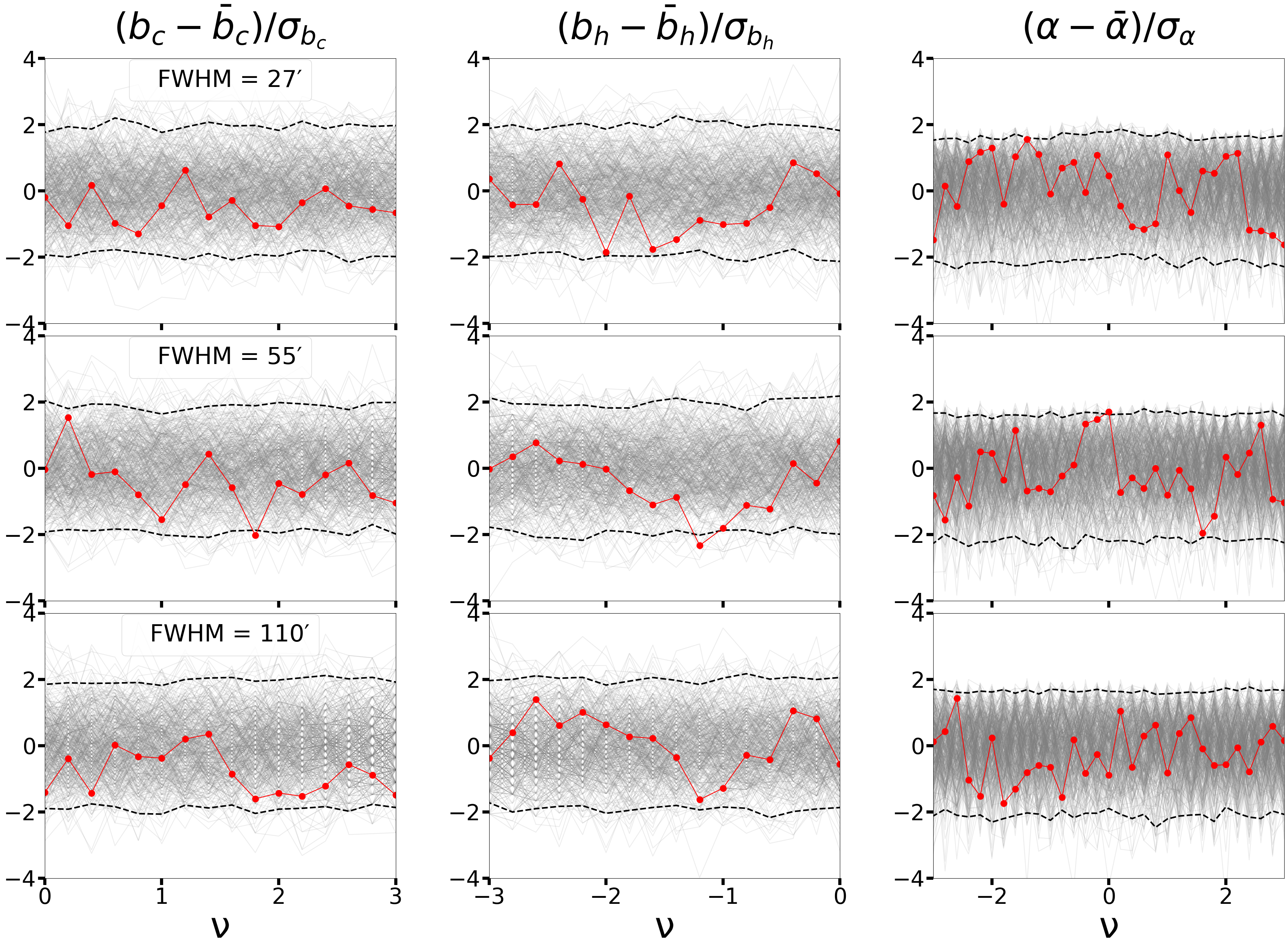}
       \vskip .2cm
2    \caption{{\bf (A)}: Deviations of the statistics $V_{0}$, $V_{1}$ and $V_{2}$ from their mean values, obtained from 400 simulations, as a function of $\nu$ for different FWHM values. The red dots (or curve) correspond to the baseline $\k$ map. The regions between the black dashed lines show the 95\% confidence interval obtained from the 400 simulations. The limits do not exactly overlap with $\pm 2$ because of the limited number of simulations.   {\bf (B)} Deviations for $b_{h}$, $b_{c}$ and $\a$ are shown. The ranges for $\nu$  are $-3< \nu < 1$ for $b_{h}$, and $-1< \nu < 3$ for $b_{c}$. The colour code is the same as in (A).} 
    \label{fig:global}
    \end{figure*}

\begin{figure*}[htp]
    \centering 
{\bf All data}\\
    \includegraphics[width=9cm]{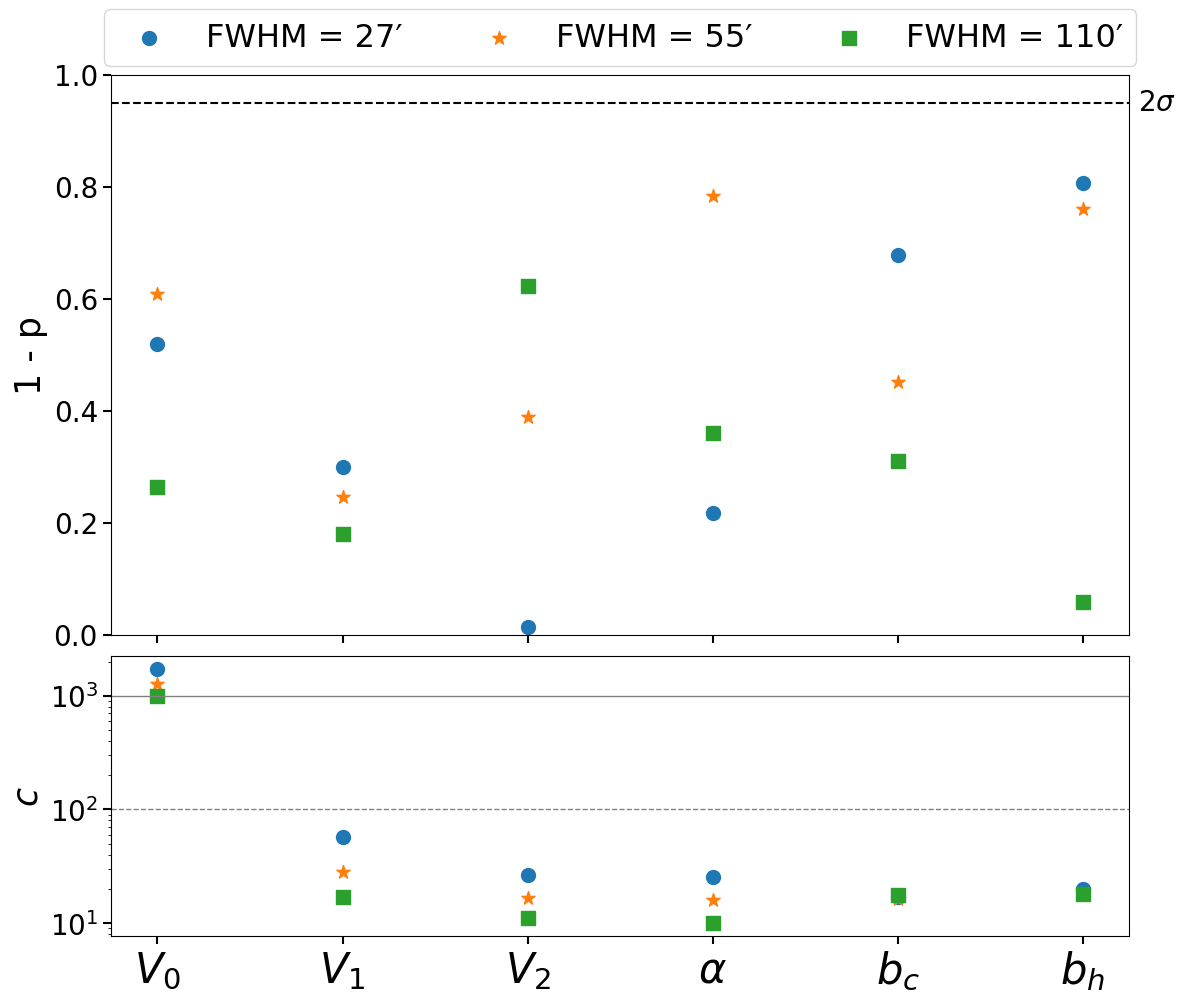}\\
      {\bf Northern hemisphere} \hskip 4.5cm {\bf Southern hemisphere}\\
        \includegraphics[width=7.5cm]{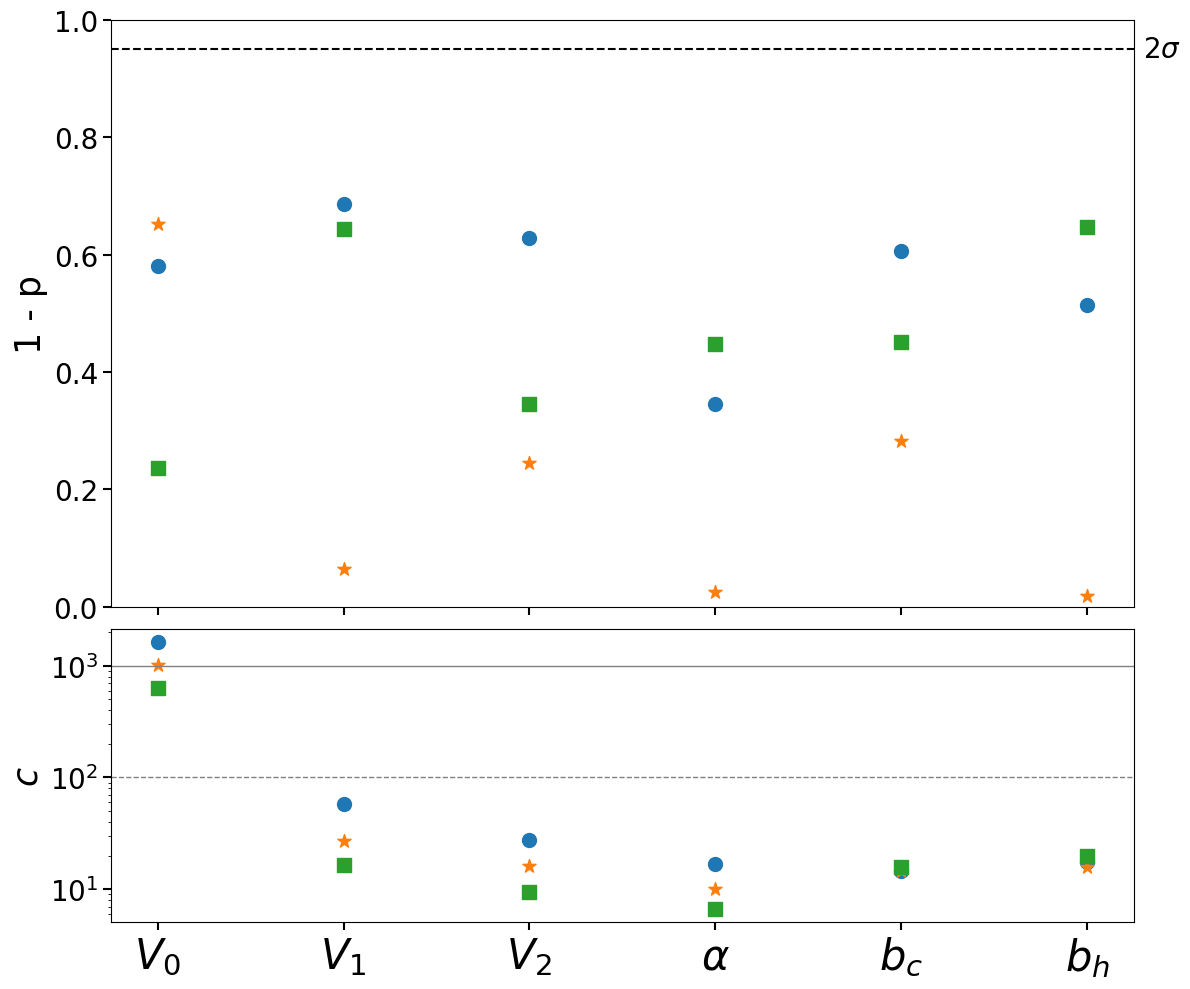} \quad
    \includegraphics[width=7.5cm]{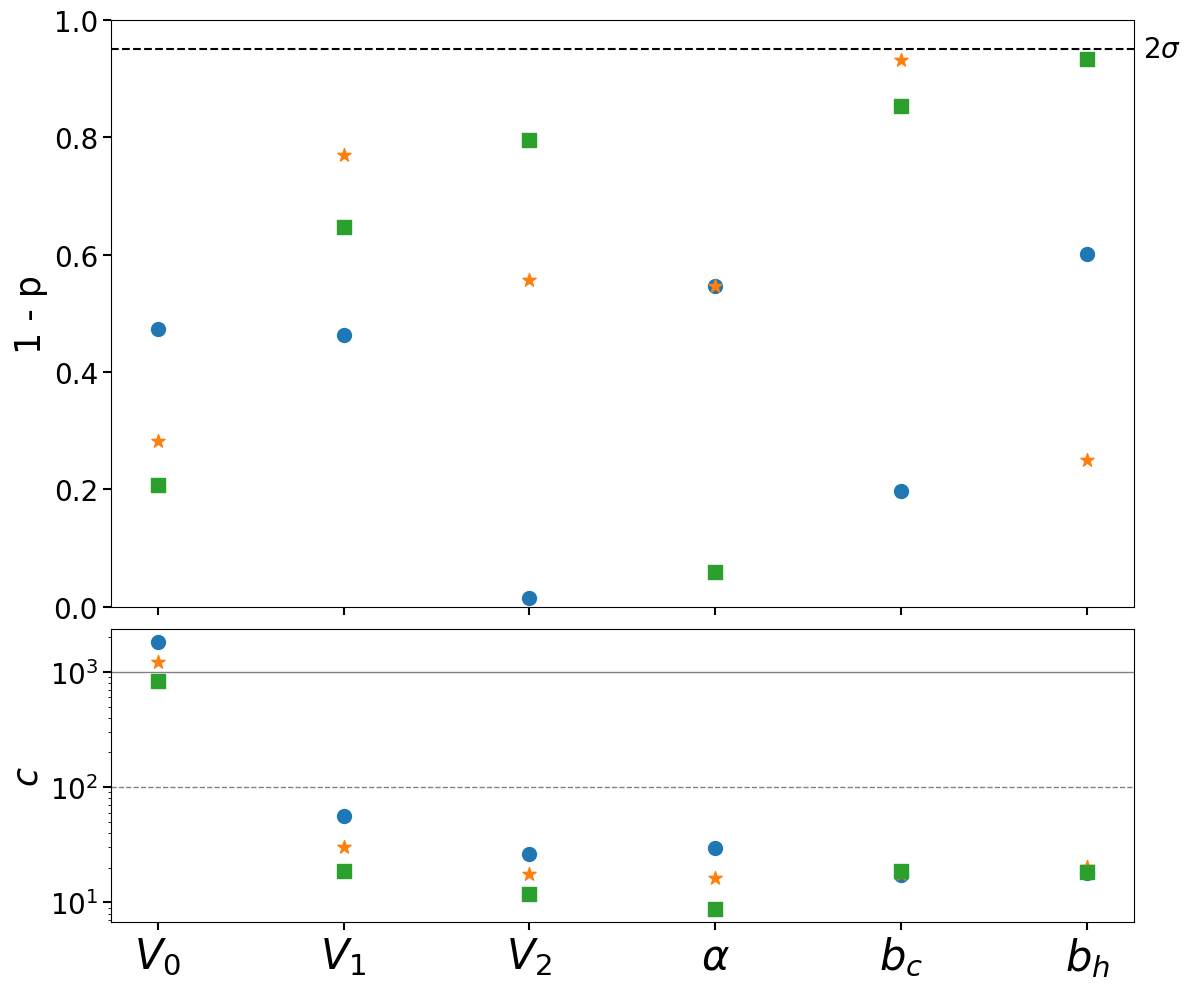}
    \caption{{\em Top}: $(1-p)$-values of various computed statistics for global analysis. The values have been assigned as per one-tailed test for $\alpha$ and two-tailed test for $V_{0}$, $V_{1}$, $V_{2}$, $b_{h}$, $b_{c}$, and have been adjusted accordingly. For reference, we have plotted $2\sigma$ line at 0.95. {\em Bottom}: $(1-p)$-values for the same quantities as the top panel, but separately computed  for the northern (left) and southern (right) hemispheres. The condition number $C$ computed from the $\Sigma_{X_iX_i}$ block of $\Sigma$  is also shown in the bottom sub-panels. The gray dashed and solid lines mark $C=100$ and 1000, respectively, for visual guide. Note that $V_0$  exhibits large values of $C$, which may lead to numerical instability in the covariance matrix. }
    \label{fig:global_patch_plot}
    \end{figure*}
    
Figure \ref{fig:global} shows the deviation $\D X _\mu(\nu_{i})$ defined by eq. (\ref{eq:Xidev}) (black dots) for each morphological statistic.  The grey shaded region shows the 95\% confidence regions obtained from the 400 simulations. We find that all statistics lie within the 95\% confidence  regions, except at very few threshold values. This indicates good agreement between the ACT observed $\k$ and the simulated $\k$ field.  

To quantify the (dis)agreement between observed data and simulations by incorporating the cross-correlations across different thresholds, we calculate the $p$-values using a one-tailed test for each statistic. The values obtained are shown in figure \ref{fig:global_patch_plot} for different FWHM 
values (denoted by different colours and shapes of the markers). The black dotted line in each panel represents the $2\sigma$ level at the 95th percentile. The methodology for computing the $p$-values is detailed in Appendix \ref{sec:a1}. All the global statistic $(1-p)$-values lie below the $2 \sigma$ level, therefore we conclude that global analysis indicates good agreement between ACT data and simulations. 

The bottom sub-panels show the corresponding condition number $C$ computed using the $\Sigma_{X_iX_i}$ block of $\Sigma$,  for each statistic $X_i$. The gray dashed and solid lines mark $C=100$ and 1000, respectively.  
Across all the plots, we observe that $V_0$ exhibits a significantly large condition number, with $C \sim 1000$, whereas the other statistics have substantially lower values, with $C < 100$. Furthermore, there is a systematic decrease in $C$ from $V_0$ to $\alpha$, which correlates with the structure of the correlation matrix shown in figure~\ref{fig:cov}.


Before analyzing smaller sky patches, we recall that the ACT DR6 reconstructed convergence simulations include non-Gaussianity from CMB lensing and the reconstruction pipeline, but not the intrinsic non-Gaussianity from nonlinear structure formation.  To assess the impact of this omission, we compare the morphological statistics of a single Agora ray-traced $N$-body convergence map \cite{Omori:2024} with 400 Gaussian simulations generated from the same fiducial convergence power spectrum at FWHM=$27',55'$ and $110'$. 
At FWHM$=27'$, 
the Agora map differs from the Gaussian ensemble at $\gtrsim 2\sigma$ significance in $V_0$, $V_1$ and $V_2$, indicating the presence of intrinsic non-Gaussianity from nonlinear structure formation. At FWHM$=55'$ and $110'$,  
the deviations fall below $<2\sigma$, indicating that the intrinsic non-Gaussianity is negligible at these angular scales and hence its omission is unlikely to affect the main conclusions of our paper.

\subsection{Analysis of local patches}
\label{sec:sec5b}

We now turn to the analysis of smaller, non-overlapping sky patches. As discussed in \cite{Goyal:2021}, the need for complementing the global analysis by patch analysis arises because any anomaly in localized sky regions
can get washed out when the morphological statistics are computed over larger regions. On the other hand, large  structures that are correlated across large angular scales can also get washed out when we probe sky regions sizes below their correlation scales. Moreover, in the global analysis we do find some thresholds where the morphological statistics show relatively large deviations beyond 2$\sigma$. Therefore, analysis in smaller patches can reveal local anomalies more sharply.

\subsubsection{Construction of patches and resolutions for the analysis}
\label{sec:sec61}

\begin{figure*}[htp]
    \centering 
    {\bf BR2} \hskip 8cm {\bf BR4}\\
    \includegraphics[width=0.9\textwidth]{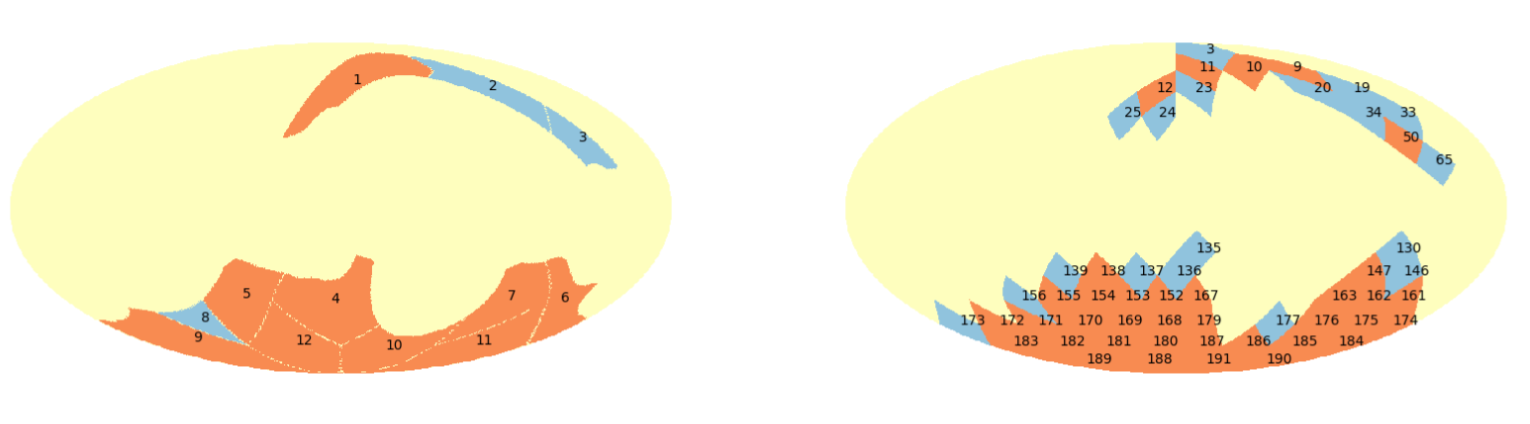}
    \caption{Sky patches for BR2 (left), and BR4 (right), numbered and shown in Galactic coordinates. In the analysis, only patches with $p_{\rm frac} \ge 0.5$ are included. These are indicated by orange colour. Blue ones have $p_{\rm frac}< 0.5$. This selection gives 9 valid patches for BR2 and 36 valid patches for BR4.}
    \label{fig:patches}
\end{figure*}
We construct sky patches in the following two ways.  
\begin{enumerate}
\item  {\em Base resolution 2} (BR2): Using $\Nside = 2$, we first divide the sky into 48 pixels, where the angular size of each pixel is roughly $29.3\,{\rm deg}^2$. Then, due to the unequal coverage of the 48 different pixels in the ACT map, we merge some regions to create patches having comparable sky fractions. The resulting patches are numbered and shown in the left panel of figure \ref{fig:patches}. Note that the numbering here does not follow \texttt{HEALPix} convention. 
For the 23\% sky fraction of ACT DR6 we get 12 patches. The blue and orange colours of the patches are explained below. 
 
\item{\em Base resolution 4} (BR4): Each of the 192 pixels obtained from $\Nside=4$ is identified as a sky patch.  The angular size of each patch is roughly $14.7\,{\rm deg}^2$. This gives 54 patches corresponding to 23\% sky fraction of ACT DR6. The patches are numbered in ring format of \texttt{HEALPix} (unlike BR2 above) and illustrated in the right panel of figure \ref{fig:patches}. The blue and orange colours for the patches are explained below. 
\end{enumerate}
Note that patches in BR4 are subsets of patches in BR2, which in turn are subsets of hemispherical patches. Thus, our  construction of patches allows for a spatially hierarchical search for anomalous behaviour.  Next, for setting the resolution of pixels within the patches, we use $\Nside =$ 512, 256 and 128, with associated smoothing scales $27', 55'$ and $110'$, respectively. These three cases are chosen keeping in mind the SNR of the observed convergence map and requiring that the patches should have good enough resolution for analysis using morphological statistics. As mentioned in section \ref{sec:s3C}, we henceforth refer to the scales probed in terms of FWHM. Varying 
FWHM 
permits a multi-scale  search for anomalous behaviour of a given sky region. 

\noindent
{\em Masking for patches}: Mask maps which have binary values, 1 for pixels inside the patch and 0 elsewhere, are prepared for each patch. Field derivatives are computed for the full data set. 
Then for computation of morphological statistics in the patches, this patch mask is applied, and only pixels where (a) the smoothed full sky mask value is $\ge 0.9$, and (b) patch mask value is one, are included. This minimizes spurious numerical errors due to mask boundaries, and maximally retains the sky fraction of each patch.  Masking is applied identically to the baseline map and simulations.

\noindent
{\em Valid patches}: We further exclude patches that have high percentage of masked regions. To quantify this, we calculate $p_{\rm frac}$ which is the ratio of the number of pixels that are not masked (valid pixels) to the total number of pixels in each patch. A patch is included in further analysis if it has $p_{\rm frac} \ge 0.5$. This gives 9 valid patches for BR2. 
For BR4, the number of valid patches is different for different $\Nside$, and we include only those patches that are common to all. This gives 36 valid patches.

\subsubsection{Methodology for identification of anomalous patches}
\label{sec:s5.2.2}
 
In the forthcoming analysis, we exclude $V_0$ because it has high condition number, as seen in section \ref{sec:sec5a}. 
 We now group  the statistics together into four sets, as follows.
\begin{enumerate}
\item $S_+\equiv \{ V_1, V_2, b_c\}$ in the threshold range  $0 \le \nu \le \nu_{\rm max}$. 
\item  $S_-\equiv \{ V_1, V_2, b_h\}$ in the threshold range  $\nu_{\rm min} \le \nu \le 0$. 
\item $\a_+$:  $\a $ in the threshold range  $0 \le \nu \le \nu_{\rm max}$. 
\item $\a_-$: $\a$ in the threshold range  $\nu_{\rm min} \le \nu \le 0$.
\end{enumerate} 
The value of $\nu_{\rm min}$ is taken to be the maximum of the minimum values of the 400 simulations of $\kappa$ for each patch. 
Likewise, $\nu_{\rm max}$ is taken to be the minimum of the maximum values of the 400 simulations of $\kappa$ for each patch.  This choice ensures that the comparison of observed data with the simulations in each patch uses only the threshold range where all simulations have field values. Both  $\nu_{\rm min}$  and $\nu_{\rm max}$  vary from patch to patch, and their magnitudes  decrease with increasing FWHM. 

The split of the threshold ranges into positive  for $S_+$ and negative for $S_-$ is `natural' for two reasons. The first is that for large scale structures, positive thresholds correspond to high density structures such as nodes, filaments and big clusters, while negative thresholds correspond to low density void regions. 
Hence, it is important to identify outliers separately for positive and negative thresholds for all statistics. 
The second reason is that $b_c$ and $b_h$ take values predominantly in the positive and negative threshold ranges, respectively (see figure \ref{fig:WF}). Hence, separating the ranges allows for taking into account the cross-correlations among the MFs and with Betti numbers. However, a caveat is that consecutive outliers 
can artificially lose significance if they are in the vicinity of $\nu=0$.  In such cases, we will extend the threshold cut beyond $\nu=0$, within reason. We will encounter one such case (patch 185 in BR4) in section~\ref{sec:s5.2.3}.  
The statistics, $\a_+$ and $\a_-$ provide information of the alignment of the structures of the excursion sets of the field, which is an independent information compared to MFs and Betti numbers. Hence, we do not cross-correlate $\a$ with the other morphological statistics.

For $S_+$ and $S_-$, we calculate $\Delta X_i$  defined by eq.  (\ref{eq:Xidev}) for each constituent statistic in each patch. Then for determining the statistical significance of the deviations, we compute $\chi^2$ given by eq. (\ref{eq:Xchi}) by averaging over a threshold range of width $D$ about central threshold values denoted by $\nu_c$. 
This average is equivalent to a {\em boxcar moving average},  with the difference that here we incorporate the covariance matrix. Here, we present results for $D=1$, meaning the averaging is over the threshold range $ [\nu_c-0.4$,  $\nu_c+0.4$], which corresponds to averaging over 1$\s$ values of the field. All the calculations are repeated for varying $D$ to check robustness of the results, though we do not include those here. Then, the $p$-values for $S_+$ and $S_-$ are calculated by comparing the computed $\chi^2$ with the corresponding scaled $F$-distribution  (as described in Appendix \ref{sec:a11}). For $\a$, we construct a distribution using 400 values from the simulations at a given threshold. If the observed value is below the median, we use the normalized percentile as the $p$-value. And, if it is above the median, we calculate the $p$-value by subtracting the normalized percentile from 1. This calculation ignores the correlation of $\a$ across thresholds, which is negligible as seen in figure~\ref{fig:cov}. 


At any given threshold $\nu_c$, if any one of the four statistics, namely, $S_+$, $S_-$, $\alpha_+$, or $\alpha_-$, exhibits a $(1 - p)$-value greater than 0.95 (corresponding to a 2$\sigma$ significance level), we initially classify the corresponding patch as an outlier.
Such outlier behavior may arise from statistical fluctuations or residual systematics, including foreground contamination or inaccurate noise modeling, which can lead to the erroneous identification of a patch as anomalous. To reduce the likelihood of such misclassifications, we subject each outlier patch to further scrutiny using the following three-step procedure.
\begin{enumerate}
\item {\bf Checking persistence of deviations}:  We consider a patch as anomalous if it exhibits outlier behavior at two or more consecutive $\nu_c$, and also  exhibit outlier behaviour at two or more FWHM values. 
This amounts to identifying anomalous patches as those that exhibit persistence of the deviations of the observed data from the expected behavior across different threshold ranges and different spatial resolutions.  We provide justification for this in Appendix~\ref{sec:corr}. 

\item {\bf Cross-checking with null maps}: The null maps (described in section \ref{sec:sec2b}) serve to test whether outlier behaviour of any patch obtained from step 1 are likely to originate from either residual foreground signals or instrument systematics. For this, we compare morphological statistics calculated for the observed and simulated null maps on sky patches, as done for the convergence maps. 

\item {\bf Estimation of false positive rate (FPR)}: Each patch identified as anomalous using steps 1 and 2 has some probability that it is anomalous due to random chance. To quantify this probability, we assign a false positive rate (FPR) value for each anomalous patch. Given the limited number of simulations (400), values of FPR below a certain limit cannot be trusted. Therefore, we conservatively set the lower bound on the false positive rate to $0.25 \%$,  implying 1 false positive in 400 simulation. The precise definition of FPR is  given in appendix \ref{sec:a2}.
\end{enumerate}

\renewcommand{\arraystretch}{1.1}

\begin{table}[h]
\caption{Anomalous patch IDs for BR2 and BR4 that are persistent in both $\nu_c$ for different FWHM. 
Note that patch 185 has outlier behaviour in the vicinity of $\nu_c=0$ spanning positive and negative sides for FWHM=$27'$.
We have included it for $S_+$ since the threshold range correlates with the outliers for the other two FWHM cases.  
The last column gives the false positive rate values. The color used for the patches is described as follows: red and orange indicates those anomalous patches having deviation $\geq 3 \s$ and $\geq 2 \s$ for atleast one FWHM, 
whereas cyan indicates deviation $\geq 2 \s$ in case of $\a$.} 
\vspace{0.1cm}
\centering
\begin{tabular}{|c|c|c|c|c|l|}
\hline
\multirow{2}{*}{\texttt{Patch size}} & \multirow{2}{*}{\texttt{Statistic}} 
& \multirow{2}{*}{\texttt{512}} & \multirow{2}{*}{\texttt{256}} & \multirow{2}{*}{\texttt{128}} & \texttt{False positive}\\
&&&&&\texttt{rate ($\%$)}\\
\hline\hline

\multirow{4}{*}{BR2} 
& $S_+$ & (\red{11}) & (\orange{11}) & \red{11} & 0.0\\ \cline{2-6}
& $S_-$ & -- & -- & -- & -- \\ \cline{2-6}
& $\alpha_+$ & -- & -- & -- & -- \\ \cline{2-6}
& $\alpha_-$ & -- & -- & -- & --\\
\hline\hline

\multirow{4}{*}{BR4} 
& \multirow{2}{*}{$S_+$}
& \orange{184} & (\red{184}) & -- & 0.09 \\ \cline{3-6}
& & \orange{185} & \red{185} & \orange{185} & 0.53 \\ \cline{2-6}

& \multirow{2}{*}{$S_-$} 
& \orange{155} & \orange{155} & -- & 1.93 \\ \cline{3-6}
& & (\red{191}) & (\red{191}) & -- & 1.39 \\ \cline{2-6}

& \multirow{2}{*}{$\alpha_+$}
& \cyan{185} & -- & -- & 1.91 \\ \cline{3-6}
& & \cyan{188} & -- & -- & 1.05 \\ \cline{2-6}

& $\alpha_-$ & -- & -- & -- & -- \\
\hline
\end{tabular}

\label{tab:nuc_and_nside}
\end{table}

\subsubsection{Anomalous patches identified}
\label{sec:s5.2.3}

The plots for $(1-p)$-values as functions of $\nu_c$ for all outlier patches are shown in figure \ref{fig:pplots} in appendix \ref{sec:a3}. They provide information of how statistically significant the deviations are. The corresponding plots for the same patches of the null map are shown in figure \ref{fig:pplots_null} in appendix \ref{sec:a4}.  In both figures, the condition numbers are also shown for all statistics, except $\a$ for which $C$ is not computed. We observe that $C$ ranges from order 10 to a few 100s, indicating that the covariance matrix inversions are sufficiently stable. We examine each outlier patch to check for persistence of the deviations and compare with the behaviour of the same patch in the null map. Then, we also compute FPR for each anomalous patch.

\begin{figure*}[htp]

\noindent
\begin{minipage}[c]{0.30\textwidth}
    \centering
    {\bf BR2}, $S_+$ 
    \quad $\longrightarrow$
\end{minipage}
\hfill
\begin{minipage}[c]{0.68\textwidth}
    \centering
    \includegraphics[width=0.60\linewidth]
    {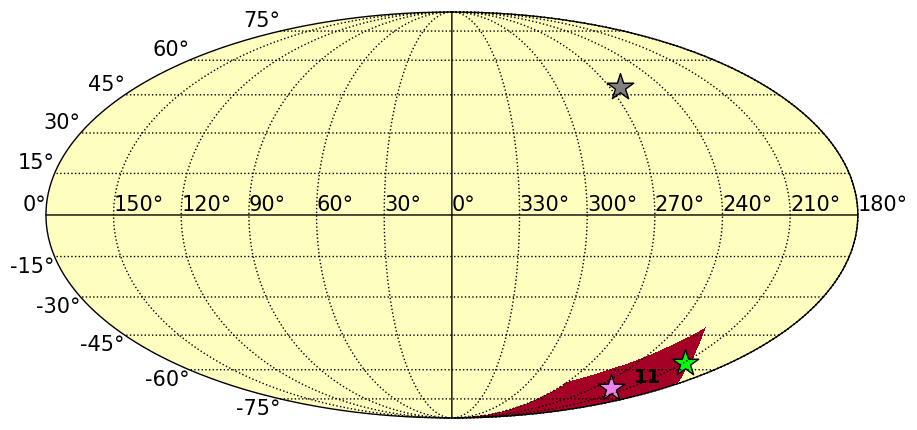}
\end{minipage}

\vspace{0.15cm}

\noindent
\begin{minipage}[c]{0.30\textwidth}
    \centering
    {\bf BR4}, $S_+$
    \quad $\longrightarrow$
\end{minipage}
\hfill
\begin{minipage}[c]{0.68\textwidth}
    \centering
    \includegraphics[width=0.60\linewidth]
    {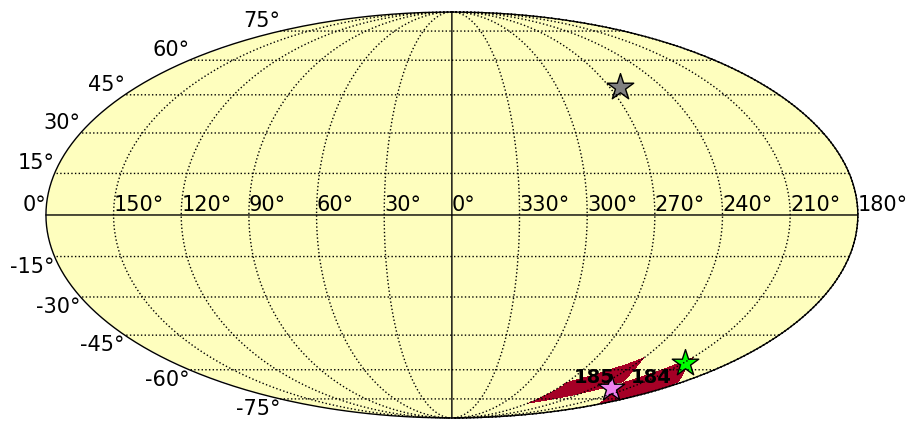}
\end{minipage}

\vspace{0.15cm}

\noindent
\begin{minipage}[c]{0.30\textwidth}
    \centering
    {\bf BR4}, $S_-$
    \quad $\longrightarrow$
\end{minipage}
\hfill
\begin{minipage}[c]{0.68\textwidth}
    \centering
    \includegraphics[width=0.60\linewidth]
    {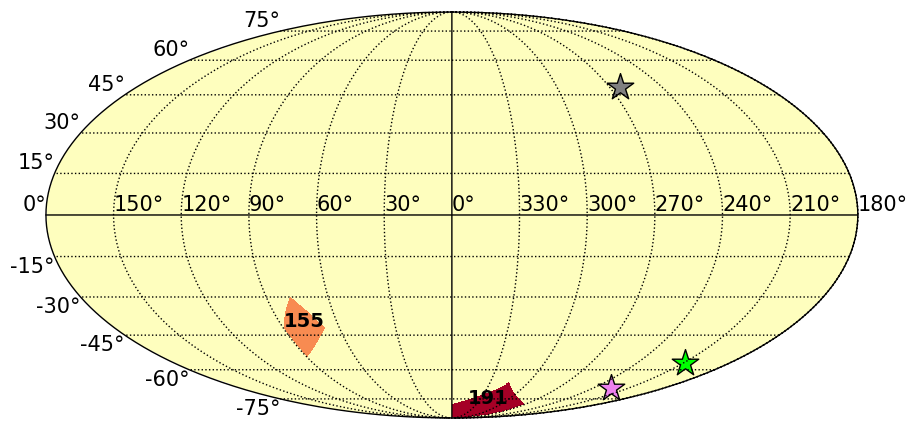}
\end{minipage}

\vspace{0.15cm}

\noindent
\begin{minipage}[c]{0.30\textwidth}
    \centering
    {\bf BR4}, $\alpha_+$
    \quad $\longrightarrow$
\end{minipage}
\hfill
\begin{minipage}[c]{0.68\textwidth}
    \centering
    \includegraphics[width=0.60\linewidth]
    {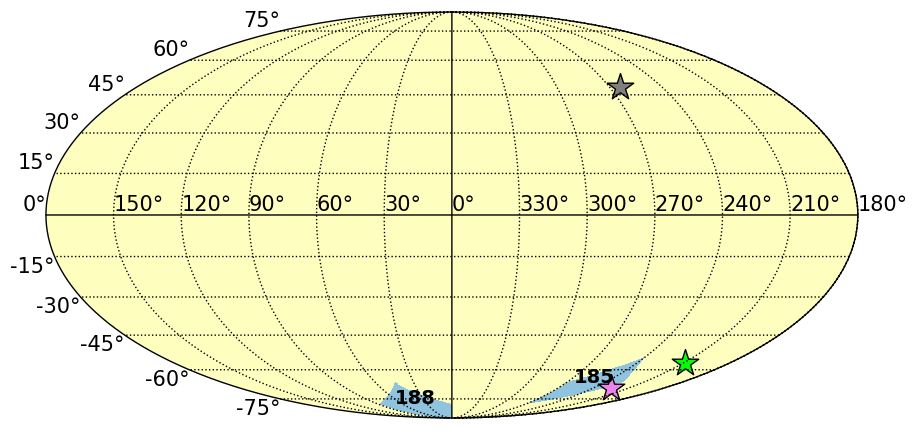}
\end{minipage}

\vspace{0.1cm}

  \caption{Anomalous patches for BR2 using $S_+$ {\bf (I)} and BR4 identified using $S_+$ {\bf (II)}, $S_-$ {\bf (III)} and $\a_+$ {\bf (IV)},   shown in Galactic coordinates and Mollweide projection. For $S_+$ and $S_-$, orange and red colours represent higher than 2$\s$ and 3$\s$ deviations, respectively. For $\a$ (IV), light blue represents 2$\sigma$ deviation with $\a$ of the observed data being lower than the median value. The patches marked here are the ones shown in table \ref{tab:nuc_and_nside}. The gray, green and violet stars represent the CMB dipole direction,   
   the Cold Spot  
   \cite{Vielva:2004} and an anomalous southern spot  
   reported in \cite{Collischon:2024}, respectively. }
    \label{fig:apatches}
\end{figure*}

The anomalous patches identified using the statistics $S_+$, $S_-$, $\alpha_+$, and $\alpha_-$, based on the persistence criteria, are listed in table~\ref{tab:nuc_and_nside}. For $\alpha$, we do not find any patch exhibiting persistent deviations across multiple FWHM 
values. Nonetheless, since $\alpha$ shows the lowest correlation of structures across different FWHM 
values (see figure~\ref{fig:corr}), we also include patches that display anomalous behavior at a single FWHM 
in the table. 
The color coding used for the patches is as follows: orange indicates anomalies at a significance level greater than $2\sigma$, and red indicates those above $3\sigma$. Round brackets denote patches that show $>2\sigma$ deviations in the corresponding null map, suggesting possible contamination. For $\alpha$, light blue indicates a deviation at the 95\% confidence level, where the observed $\alpha$ value is lower than the median from simulations, indicating a tendency toward aligned structures. 
The sky locations of all anomalous patches identified using $S_+$, $S_-$, and $\alpha$ are shown in figure~\ref{fig:apatches}, in Galactic coordinates using a Mollweide projection. The color coding of the patches is consistent with that in table~\ref{tab:nuc_and_nside}. The grey, green, and violet stars mark the directions of the CMB dipole, the Cold Spot (CS) \cite{Vielva:2004} and an anomalous southern spot (SS) reported in \cite{Collischon:2024}, respectively.

\begin{figure*}[t]
\centering

\includegraphics[width=.5\textwidth]{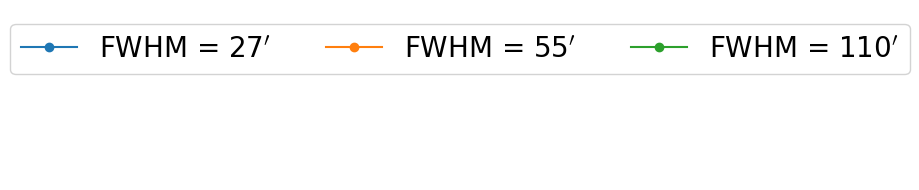}\\[-0.6cm]

\begin{minipage}{0.95\textwidth}
\centering
\begin{tabular}{@{}c@{\hspace{2.7cm}}c@{\hspace{1.7cm}}c@{\hspace{1.7cm}}c@{}}
BR2, $S_+,S_-$ & BR4, $S_+,S_-$ & BR4, $S_+,S_-$ & BR4, $S_+,S_-$ \\
\end{tabular}
\end{minipage}\\[0.2cm]

\begin{minipage}{0.95\textwidth}
\centering
\includegraphics[width=.25\textwidth]{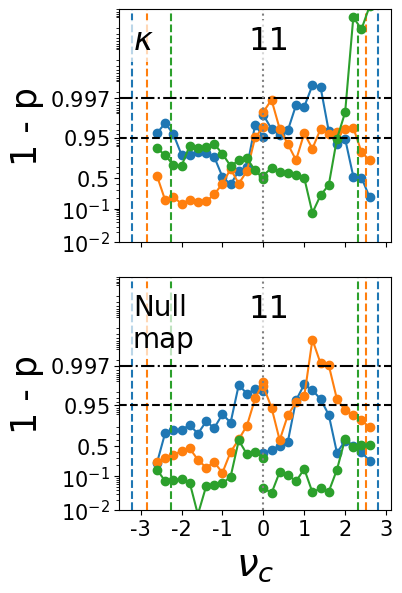}\hspace{0.3cm}
\includegraphics[width=.625\textwidth]{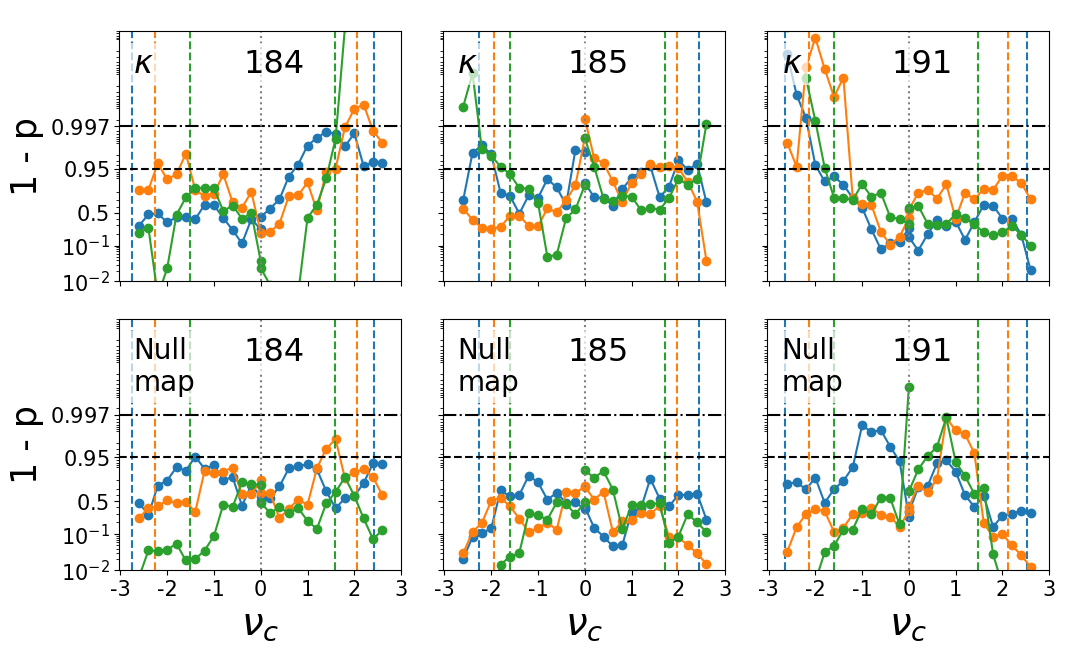}
\end{minipage}\\[0.3cm]

\begin{minipage}{0.95\textwidth}
\centering
\begin{tabular}{@{}c@{\hspace{1.7cm}}c@{\hspace{1.7cm}}c@{}}
BR4, $S_+,S_-$ & BR4, $\alpha_+$ & BR4, $\alpha_+$ \\
\end{tabular}
\end{minipage}\\[0.2cm]

\begin{minipage}{0.95\textwidth}
\centering
\includegraphics[width=.625\textwidth]{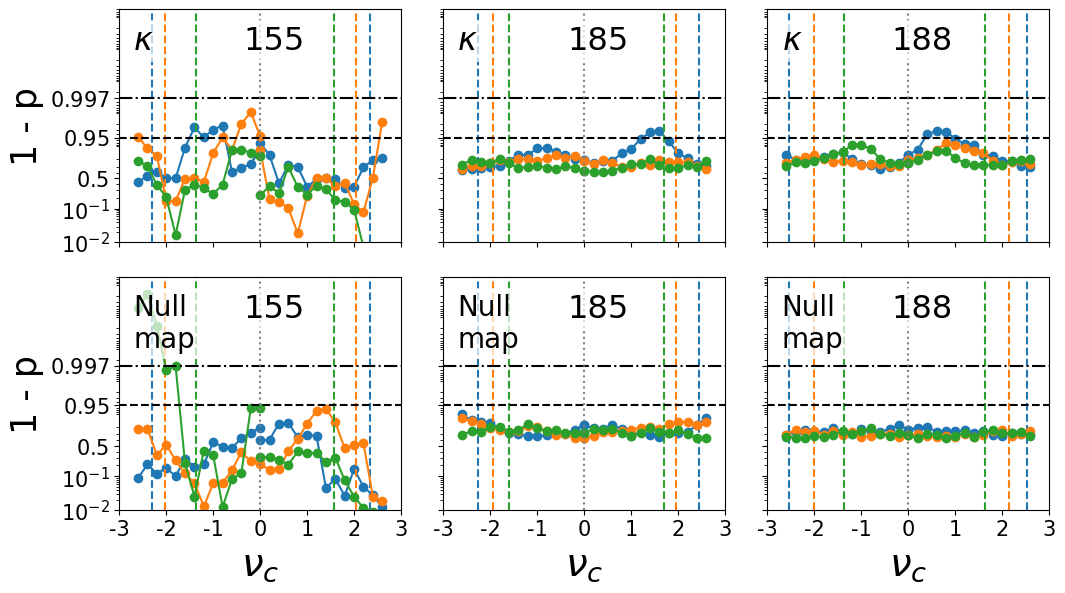}
\end{minipage}

\caption{
Comparison of $(1-p)$-values between the observed $\kappa$ and null map for the patches identified as anomalous in table~\ref{tab:nuc_and_nside}. 
See also figure~\ref{fig:apatches}.
}
\label{fig:f8}
\end{figure*}

We now discuss each anomalous patch one by one (not in numerical order.
\begin{itemize}
\item[$\bullet$] {\em BR2, patch 11}: This region is found to be strongly anomalous exhibiting  persistent $> 99 \%$ confidence level deviations 
$\nu_c$ and all three FWHM 
values for $S_+$. The estimated FPR value is 0 ($\leq 0.25 \%$). It is  particularly interesting because it contains both the CS and SS anomalous regions (see figure~\ref{fig:apatches}). A comparison of the $(1-p)$-values between the $\kappa$ and null map is shown in the top left panel of figure~\ref{fig:f8}. We observe that both the maps show persistent outliers at FWHM=$27'$ and $55'$, 
particularly for $S_+$. This suggests that the observed $\kappa$ map contains residual foreground contamination and/or instrument systematics and noise in this region. At FWHM=$110'$, 
the null map does not show any outliers, while $\kappa$ shows highly significant outliers  consecutively at $\nu_c=2,2.25$. Given that the signal to (reconstruction) noise in $\kappa$ becomes of order one at this FWHM, 
the anomalous nature of this patch suggests anomalous nature of the underlying matter distribution in sub regions where the matter density is $\gtrsim 2\sigma$ ($\nu_c \gtrsim 2$). This patch includes as subsets anomalous patches 184 and 185 in BR4, and closer examination of these can further help in isolating the cause(s) of anomaly.
\item[$\bullet$] {\em BR4, patch 184}: This patch is found to be strongly anomalous exhibiting  persistent $ > 99 \%$ confidence level deviations both in $\nu_c$ and FWHM=$27'$ and $55'$, 
for $S_+$ towards $\nu_c\gtrsim 1$. The estimated FPR value is 0.09 ($\leq 0.25 \%$). In comparison  the null map shows anomaly for only FWHM=$55'$ for $S_+$ (see first row, second from left panel, of figure~\ref{fig:f8}). This suggests residual foreground contamination and/or instrument systematics and noise. It is also worth mentioning that for FWHM=$110'$, even though our strict criteria excludes this patch from being anomalous,  the trend for $(1-p)$-values towards $\nu_c>1$ is similar to what was seen for patch 11 of BR2. Also, note that the CS and SS anomalous regions lie at the edges of this patch. 
\item[$\bullet$] {\em BR4, patch 185}:  This patch is found to be strongly anomalous, exhibiting  persistent $> 99 \%$ confidence level deviations both in $\nu_c$ and all three FWHM  
values, for $S_+$ towards $\nu_c\gtrsim 1$. Note that for FWHM=$27'$  
the anomalous nature extends into negative threshold in the vicinity of $\nu_c=0$. We have included it $S_+$ due to the contiguity, as mentioned in the previous subsection. The estimated FPR value is 0.53. In comparison, the null map does not show anomaly (see first row, third from left panel, of figure~\ref{fig:f8}). 
This suggests that this region does not contain residual foreground contamination and/or instrument systematics and noise. Also, note that the SS region lies at the edge of this patch.  We also observe mild anomaly, with $\alpha$ lower than the median value from simulations (implying anisotropy), for FWHM=$27'$  
for $\a_+$. 
\item[$\bullet$] {\em BR4, patch 191}: This patch is found to be strongly anomalous, exhibiting  persistent $\approx 99 \%$ confidence level deviations both in $\nu_c$ and FWHM=$55'$ and $55'$, 
for $S_-$ towards $\nu_c\lesssim -1$. The corresponding region in  the null map also shows anomaly for both these FWHM 
values, but for $\nu_c> -1$ (see first row, second from left panel, of figure~\ref{fig:f8}). These results suggest that, while this patch likey contains residual foreground contamination and/or instrument systematics and noise, the underlying matter distribution likely shows anomaly in regions where the density is $\lesssim -1\sigma$.
\item[$\bullet$] {\em BR4, patch 155}: This patch is found to be mildly anomalous, exhibiting  persistent $\approx 99 \%$ confidence level deviations both in $\nu_c$, and FWHM=$27'$ and $55'$,   
for $S_-$ towards $\nu_c\lesssim -1$. The corresponding region in the null map does not show anomaly. 
\item[$\bullet$] {\em BR4, patch 188}: This patch is found to be mildly anomalous, with $\alpha$ lower than the median value from simulations (implying anisotropy), for FWHM=$27'$ 
for $\a_+$. The corresponding region in the null map does not show anomaly. 
\end{itemize}

Comparing our results with figure 6 of~\cite{Goyal:2021} using the $\alpha$ statistic, we note that the majority of the patches flagged as anomalous in that work lie outside the ACT footprint, thereby limiting the scope of a direct comparison. However, for sky regions common to both the ACT and Planck datasets, we find no overlapping anomalous patches. 
Since the method employed here includes an additional boxcar averaging step over one-sigma threshold range, rendering it more conservative, we repeat the anomaly identification using the same criteria as in~\cite{Goyal:2021} to enable a fair comparison. From this analysis, we identify patches 175 (located near the Cold Spot) and 191 as anomalous at the 95\% confidence level, both exhibiting $\alpha$ values lower than the median. These results are presented in Appendix~\ref{sec:a5}.
Under this less stringent approach, all but one of the anomalous patches identified at each FWHM show $\alpha$ values below the median simulation values, suggesting a preference for directionally aligned structures. Patches 185 and 188 continue to appear mildly anomalous in the present analysis.

\section{Conclusion and discussion}
\label{sec:sec7}

 In this work, we have tested the standard model of cosmology using the observed ACT convergence map and comparing it with the simulations provided by the ACT team. Our approach utilizes a suite of morphological statistics which measure the geometry and topology of excursion sets of smooth random fields. We analyze the data in three regimes; the global (masked) sky, galactic hemispheres and small sky patches. One can consider that for each of these measurements we are studying a different aspect of the standard model. We also vary the smoothing scales  
 so as to  probe morphology across a range of angular scales without assuming a preferred scale {\em a priori}. As FWHM increases, 
 the Minkowski functionals become increasingly sensitive to large-scale features while suppressing small-scale structure. Consequently, anomalies that disappear at higher FWHM 
 are likely associated with small-scale contaminants, whereas those that persist across multiple smoothing scales 
 provide stronger evidence for a genuine, scale-independent signal.

The global analysis serves as a consistency test for the ACT simulations (mock data) under the assumption that the Universe is statistically isotropic and homogeneous -- an all-sky average of scalar statistics cannot uniquely determine if the data contains an anisotropic signal.  The agreement found in Section \ref{sec:sec5a} implies that the ACT data is consistent with the simulated $\Lambda$CDM expectations, indicating no significant departure from the standard cosmological model at large scales.

The hemispherical analysis then provides a test for the presence of anomalous dipolar behavior in the density field statistics. Note that this is different from dipole searches such as \cite{Secrest:2021,Secrest:2022}, which are a direct measurement of the dipole modulation of number density in a flux-limited sample. Here, we are considering the possibility that the statistical properties of large-scale structure differ in the northern and southern sky. Such a test is rational, given that the CMB itself exhibits a power asymmetry in the temperature field. In this work, we find no statistically significant differences between the two hemispheres. However, the data only cover a modest fraction of the total sky ($f_{\rm sky} = 0.23$) and therefore, the analysis should be repeated using larger datasets to further strengthen these conclusions. 

Finally, the patch-wise analysis serves as a targeted search for localized anomalous features in the data. This approach is well-motivated, given the presence of irregular local features in the CMB temperature map, such as the well-known Cold Spot.\footnote{The Cold Spot is a much smaller-scale feature compared to the patches studied here, but its presence motivates the search for other potential anomalies.} 
In our analysis, we identify a few atypical patches where the observed data differ from the simulations at a statistical significance exceeding 99\% confidence level. Notably, we identify a region  - specifically patch 11 in BR2 (that include patches 184 and 185 in BR4) - as anomalous in $S_{+}$ (see figure~\ref{fig:apatches}). This region encompasses both the Cold Spot and the anomalous southern spot, represented by the green and violet stars, respectively, in figure~\ref{fig:apatches}. The fact that this overlap is found using the convergence map, rather than the temperature data, makes these regions worthy of further investigation. Further analysis using the null map suggests that part of the anomalous behavior may arise from residual foreground contamination, instrumental systematics, or reconstruction noise. Additionally, we identify patches 188 and 191 near the Southern Galactic Pole, and patch 155 in the southern mid-latitude region, which exhibit irregular behavior in one or more of the statistical measures used in this study. A detailed investigation of the CMB data in these regions is deferred to future work. To understand the physical origin of the atypical behaviour, further study is needed,  considering possible influences from residual foregrounds, instrumental systematics, reconstruction noise, and insights from independent large-scale structure surveys. 
We note that the preponderance of unusual regions that have been found in our analysis are in the southern sky, but this is unsurprising given that the majority of the ACT data coverage lies in the south. 

The methodology employed in this study to assess the statistical significance of features in the data using morphological and topological summary statistics builds upon earlier work by two of the present authors. In~\cite{Goyal:2021}, a similar analysis was carried out on the Planck convergence map, using only the $\alpha$ statistic, and a significantly larger number of anomalous patches were reported. 
The current study extends and improves upon that work in several key ways. First, we utilize ACT data, which offers a higher signal-to-noise ratio, thereby enabling a more accurate extraction of morphological statistics. Second, we refine the statistical framework to more robustly quantify the significance of observed deviations. In particular, the criteria used here to classify a patch as nonstandard are substantially more stringent than those applied in the earlier analysis. 
Based on this improved methodology, we conclude that many of the anomalous patches identified in~\cite{Goyal:2021} were likely the result of residual foreground contamination, instrumental systematics, or reconstruction noise present in the Planck convergence map.

In the future, our analysis will be expanded further as more data becomes available, which will allow us to search for large scale features with greater statistical power. Increasing the number of mock realisations will allow us to more robustly reconstruct the covariance matrix of the summary statistics. It will also be of interest to apply the methodology to other cosmological datasets such as large scale structure maps.  Ref. \cite{Vanessa:2018} found that including the non-Gaussianity from nonlinear structure formation in convergence map reconstruction induces a percent-level bias in temperature-based power spectrum measurements. Its impact on morphological statistics globally as well as in smaller sky regions, particularly at resolutions above FWHM=$27'$, remains to be quantified using simulation suites such as Agora \cite{Omori:2024}.

\begin{acknowledgments}
We acknowledge the use of the \texttt{Nova} cluster at the Indian Institute of  Astrophysics, Bangalore. We have used \texttt{HEALPIX}\footnote{\protect\url{http://healpix.sourceforge.net/}}
~\cite{Gorski:2005}, \texttt{Healpy}~\cite{Zonca:2019}, \texttt{Matplotlib}~\cite{Hunter:2007}. FR is supported by the NASA award 80NSSC25K7506. FR acknowledges helpful discussions with Cristian Vargas. 
PG is supported by KIAS Individual Grant (PG088101) at Korea Institute for Advanced Study (KIAS). SA is supported by an appointment to the JRG Program at the APCTP through the Science and Technology Promotion Fund and Lottery Fund of the Korean Government, and was also supported by the Korean Local Governments in Gyeongsangbuk-do Province and Pohang City. SA also acknowledges support from the NRF of Korea (Grant No. NRF-2022R1F1A1061590) funded by the Korean Government (MSIT).  CBP is supported by the KIAS Individual Grant QP016901 at KIAS and by the National Research Foundation of Korea (NRF) grant funded by the Korean government (MSIT; RS-2024-00360385).
\end{acknowledgments}

\section*{Data Availability}
 
The data that support the findings of this article are openly available \cite{Madhavacheril:2021, Omori:2024}.

\appendix
\section{Quantifying uncertainties}
\label{sec:a1}

For a random vector $\mathbf{X} = (x_{1}, x_{2},.... x_{n})$, sampled from a multivariate Gaussian distribution with mean vector $\mathbf{\mu}$ and arbitrary covariance matrix $\mathbf{\Sigma}$, the log-likelihood function is given by 
\begin{eqnarray}
    \chi^{2} (n) = (\mathbf{X} - \mathbf{\mu})^{T} \mathbf{\Sigma^{-1}} (\mathbf{X} - \mathbf{\mu}), 
    \label{eqn:chisqr_def}
\end{eqnarray}
which follows a chi-squared distribution for $n$ independent degree of freedom.

\subsection{$p$-value calculation using the $F$-distribution}
\label{sec:a11}
When dealing with datasets with finite samples, we use estimators to calculate the mean  given by,
\begin{eqnarray}
    \mathbf{\hat{\mu}} = \frac{1}{N}\sum_{i=1}^{N}{\mathbf{X}}^{\rm sim}_{i},
    \label{eq:mean_num}
\end{eqnarray}
and the covariance matrix, given by,
\begin{eqnarray}
\mathbf{\hat{\S}} &=& \langle(\mathbf{X}^{\rm sim}_{i} - {\mathbf{\hat{\mu}}})({\mathbf{X}^{\rm sim}_{i}} - {\mathbf{\hat{\mu}}})^T\rangle \nn \\
&=& \frac{1}{N - 1} \sum_{i=1}^{N} (\mathbf{X}^{\rm sim}_{i} - {\mathbf{\hat{\mu}}})({\mathbf{X}^{\rm sim}_{i}} - {\mathbf{\hat{\mu}}})^T, 
\label{eq:cov_num}
\end{eqnarray}
where $\mathbf{X}^{\rm sim}_i$ is the vector corresponding to the i'th simulation, and $N$ is the number of simulations. Consequently, we can define a $\hat{\chi}^2$ variable, as,
\begin{eqnarray}
    \hat{\chi}^2 = (\mathbf{X}^{\rm obs} - \hat{\mathbf{\mu}})^T \ \hat{\S}^{-1} (\mathbf{X}^{\rm obs} - \hat{\mathbf{\mu}}).
    \label{eq:chisqr_num}
\end{eqnarray}
For our purposes, most vectors have a size ($n$) comparable to the number of simulations ($N$). In such cases, the joint likelihood no longer follows a multivariate Gaussian distribution, but follows a multivariate $t$-distribution, \cite{Sellentin:2017} which in turn means that the $\hat{\chi}^2$ variable as defined in eq. (\ref{eq:chisqr_num}) would follow an $F$ distribution, i.e., 
\begin{eqnarray}
    \hat{\chi}^2 (n, N) \sim \frac{(N-1)n}{(N - n)} \mathbf{F}_{n, N-n},
    \label{eq:sF}
\end{eqnarray}
where $\mathbf{F}_{n, N-n}$ is the $F$-distribution with $n$ and $N-n$ degrees of freedom. 

This formalism allows us to define a simple $\chi^2$-type test (similar to the one defined in \cite{Ducout:2013}) for all statistics that are modeled as Gaussian ($V_{0}$, $V_{1}$, $V_{2}$, $b_{\rm c}$ and $b_{\rm h}$). We can calculate $p$-values by comparing the $\hat{\chi}^2$ value with the corresponding scaled $F$-distribution $\mathbf{F}_{n, N-n}$. A $p$-value of 0.05 is used as the standard, and a one-tailed test is performed, which thereby corresponds to the 95th percentile, or $2\s$ confidence interval.

\begin{figure}[htp]
    \centering 

    \includegraphics[width=0.95\linewidth]{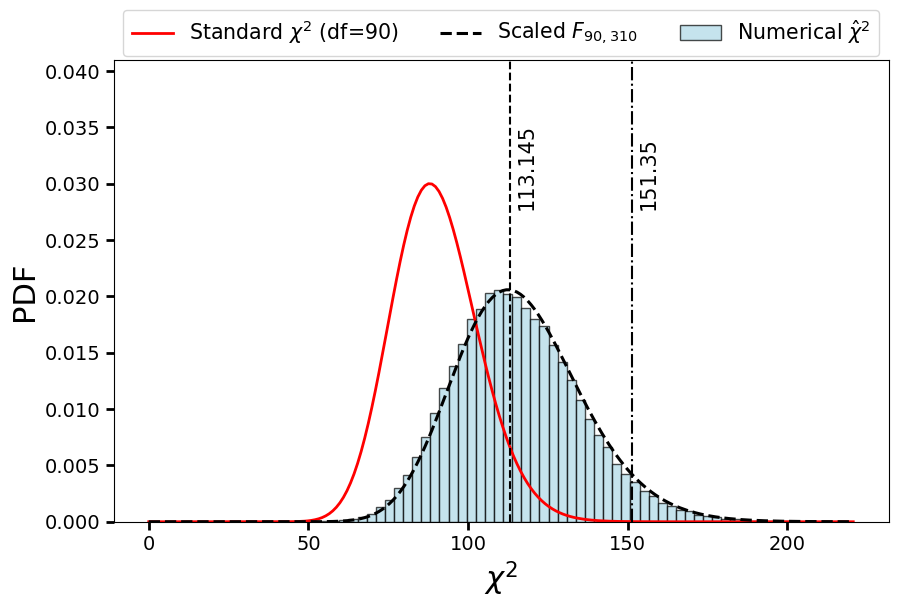}

    \vspace{0.5cm}

    \includegraphics[width=0.97\linewidth]{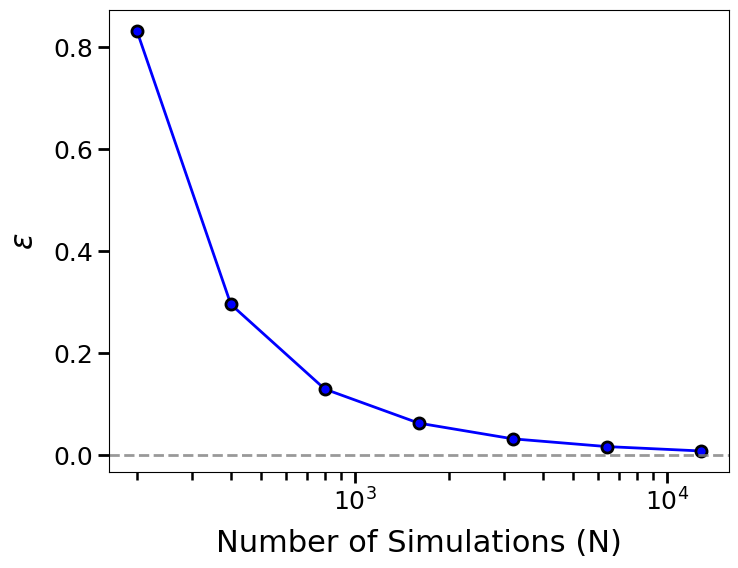}

    \caption{
    {\em Top}: Plot of PDF of $\hat{\chi}^2(90,400)$ distribution for 400 simulations (blue histogram) and standard $\chi^{2}(90)$ distribution (red curve) for 90 degrees of freedom.
    The dashed curve corresponds to the scaled $F$-distribution curve given by Eq.~(\ref{eq:sF}).
    The dashed and dotted–dashed lines correspond to the $95^{\rm th}$ percentile values for the $\chi^2$ and $\hat{\chi}^2$ distributions, respectively.
    {\em Bottom}: Plot of $\epsilon$ as a function of the number of simulations. As the number of simulations increases, $\hat{\chi}^{2}(90,N)$ converges to the standard $\chi^{2}(90)$ distribution.
    }
    \label{fig:pseudo_chisqr}
\end{figure}

We have verified this numerically by sampling, independently, a single gaussian random vector $\mathbf{X} = (x_{1}, x_{2},.... x_{n})$, to act as the observed value, and a collection of $N$ random vectors, to act as simulation data, which is then used to compute the $\hat{\chi}^2$ variable using equations \ref{eq:mean_num}, \ref{eq:cov_num} and \ref{eq:chisqr_num}. Repeating this process for $10^{5}$ realizations, we numerically estimate the distribution of the $\hat{\chi}^2(n,N)$ variable for $N$ simulations having $n$ thresholds (figure \ref{fig:pseudo_chisqr} (right) shows the distribution for $n = 90,\ N=400$).

For a given $p$-value, the deviation of the $\hat{\chi}^2(n,N)$ variable from the corresponding standard chi-squared value $\chi^2(n)$ can be quantified as,
\begin{eqnarray}
    \e = \frac{\hat{\chi}^2(n,N) - \chi^2(n)}{\chi^2(n)}.
\end{eqnarray}
As we increase the number of simulations, $\e$ approaches 0, causing the $\hat{\chi}^2$ distribution (and thereby the corresponding scaled $\mathbf{F}_{n, N-n}$ distribution) to approach the standard chi-squared distribution $ \chi^2(n)$ (figure \ref{fig:pseudo_chisqr} (left)).

\subsection{$\hat{\mathcal{M}}^2$ method}
\label{sec:a12}

The formalism discussed so far assumes that the random vectors $\mathbf{X}$ follow a multivariate gaussian distribution, making it unsuitable for Beta-distributed statistics, such as $\alpha$. However, since the covariance matrix of our dataset is close to the identity matrix (see figure \ref{fig:cov}), we can neglect correlations across thresholds. 

For our analysis, we fit a Beta distribution \( B_{\nu}(x; a_{\nu}, b_{\nu}) \), with parameters $(a_{\nu}, b_{\nu})$, to simulation data at each threshold $\nu$, and generate a sample vectors as $\mathbf{X} = (x_{1}, x_{2},.... x_{n})$, where each $x_{i}$ is sampled from the corresponding beta distribution \( B_{i}(x; a_{i}, b_{i}) \). We sample a single random vector $\mathbf{X}$ to act as the observed value, and a collection of N vectors to act as simulation data. The mean ($\hat\mu$) and covariance matrix $\mathbf{\hat{\S}}$ are calculated from equations \ref{eq:mean_num} and \ref{eq:chisqr_num}. The samples random vectors are all uncorrelated, which is the underlying assumption for this method.

We now define a modified chi-squared variable, as
\[
\hat{\mathcal{M}}^2(n, N) = (\mathbf{X}^{\rm obs} - \hat{\mu})^T \hat\Sigma^{-1}(\mathbf{X}^{\rm obs} - \hat{\mu}).
\]
This definition is identical to eq. (\ref{eq:chisqr_num}), but the constructed variable $\hat{\mathcal{M}}^2$ doesn't follow any standard distribution, hence we simulate it numerically.

Iterating over \( 10^5 \) realizations and fixing the $p$-value as 0.05, we compute the $95^{\rm th}$  percentile of $\hat{\mathcal{M}}^2(n, N)$ for one-tailed test, corresponding to a $2\sigma$ confidence interval. Unlike the $\hat{\chi}^2$ variable, the distribution of $\hat{\mathcal{M}}^2$ variable is unique for each dataset, as the \( B_{\nu}(x; a_{\nu}, b_{\nu}) \) for a particular threshold $\nu$ is different for different datasets.

\section{Quantification of correlations of structures across thresholds and different smoothing FWHM values} 
\label{sec:corr}

\begin{figure}[htp]
    \centering
    \includegraphics[width=1.00\linewidth]{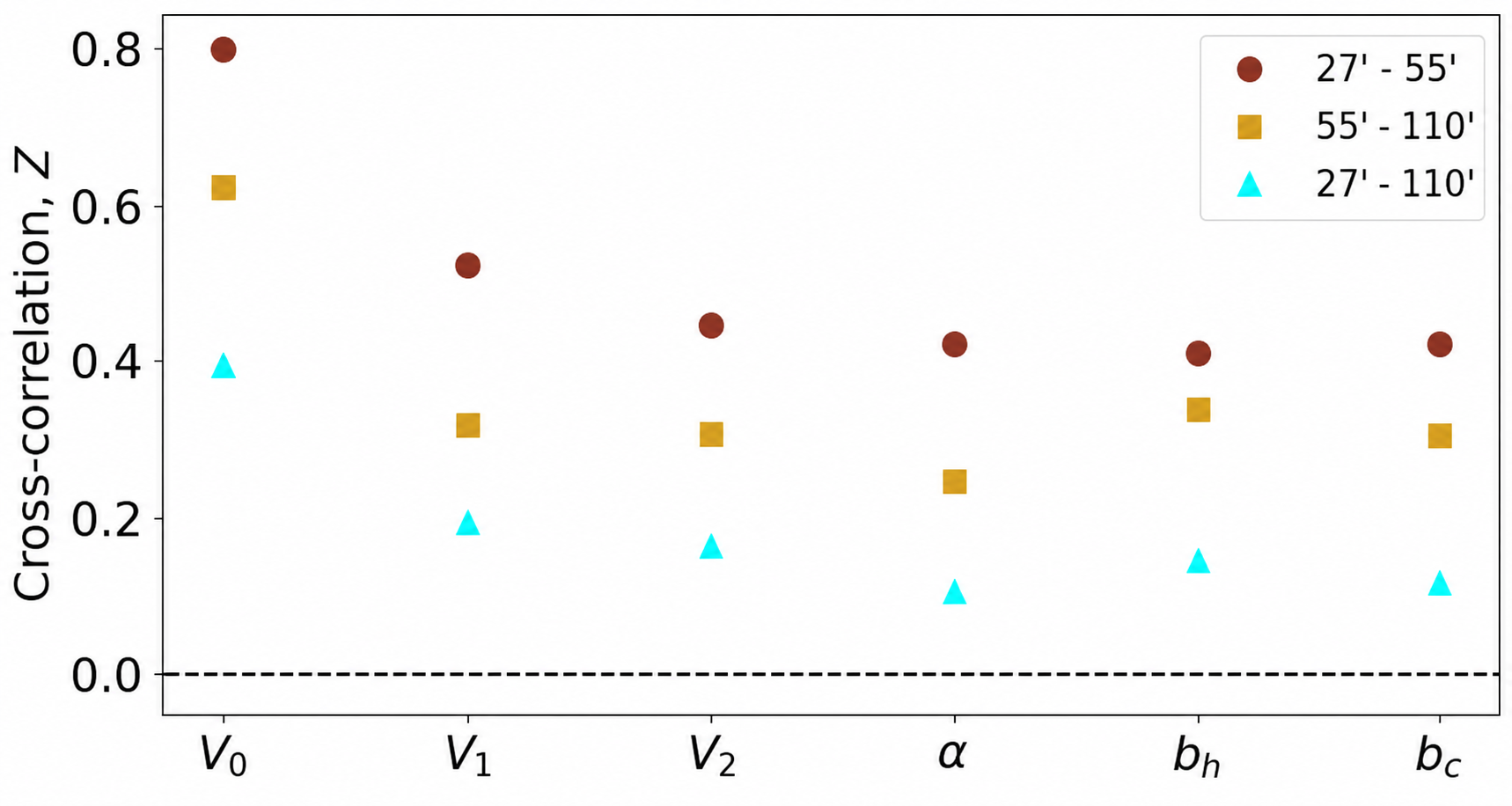}
    \caption{Cross-correlation, $Z$,  defined by eq. (\ref{eq:z}) for each pair of FWHM 
    values. } 
    \label{fig:corr}
\end{figure}

In section~\ref{sec:s5.2.2}, we use the criteria of persistence of deviations of the statistics across thresholds and FWHM 
values. This criteria is based on the expectation that (truly anomalous) structures must be correlated across different (neighbouring) FWHM  
values. Here, we justify this expectation by quantifying the cross-correlations of structures, denoted by $Z$,  for each pair of FWHM 
values, as,

\be
\begin{aligned}
Z(k,n_1,n_2) ={}&
\bigg[
\bigg\langle \sum_i X_k^{n_1}(i)\, X_k^{n_2}(i) \bigg\rangle
- \\
\bigg\langle \sum_i X_k^{n_1}(i) \bigg\rangle
  \bigg\langle \sum_i X_k^{n_2}(i) \bigg\rangle
\bigg] 
&\bigg/ \left( \sigma_{X_k^{n_1}}\, \sigma_{X_k^{n_2}} \right),
\end{aligned}
\label{eq:z}
\ee
where $k$ indexes the statistics, $i$ runs over the thresholds, $n_1,n_2$ refer to either FWHM=$27', 55'$ and $110'$ 
and $\langle.\rangle$ denotes average over 400 simulations of $\kappa$. The denominator is the product of the corresponding standard deviations.

Figure~\ref{fig:corr} presents the values of $Z$ for all the statistics, evaluated for three pairs of $(n_1, n_2)$. What is shown are $Z$ for the full sky data (not patches).  As expected, we find positive correlations across all statistics. Notably, the pair $(512, 128)$ exhibits lower correlation values compared to the other two pairs, which involve consecutive FWHM 
values. Additionally, the overall trend of $Z$ across the different statistics generally follows that of the condition number $C$ shown in figure~\ref{fig:global_patch_plot}.

\section{False positive rate} 
\label{sec:a2}

For each base resolution, let $P_j$ denote each valid patch, where $j=1,\ldots,N_P$, with $N_P$ being the number of valid patches. Let $P_a$ denote an anomalous patch. Note the nature of anomaly of $P_a$, meaning at how many $\nu_c$'s we get $(1-p)>0.95$ and for which FWHM values. 
\begin{itemize}
\item   
For each patch $P_j$, we first wish to assign a false positive weight, $\omega_i$, to each  simulation, indexed by $i$. For this purpose, let us choose the $i$th simulation as mock observed data and use the remaining 399 to compute the covariance matrix and calculate $(1-p)$ for all FWHM values. 
Next, by examining the $(1-p)$-values we check if $P_j$ contains anomaly similar to $P_a$. If it does,  then we assign $\omega_i=1$, else $\omega_i=0$. This process is repeated for all simulations $i=1$ to 400. Then we calculate the average, $\bar\omega_j$, for each $P_j$,  
\be
\bar\omega_j = \frac{1}{400}\sum_{i=1}^{400} \omega_i.
\ee
Note that the set of values $\{\o_i\}$ will differ from patch to patch. 
\item 
Then, we define the FPR for anomalous patch $P_a$  to be 
\be
FPR = \frac{1}{N_P}\sum_{j=1}^{N_P} \bar\omega_j.\ee
\end{itemize}

This method assumes that the FPR for a particular nature of anomaly is the same for all patches $P_j$, which allows us to average the FPR across $P_j$.

\section{Plots of $(1-p)$-values for the outlier patches} 
\label{sec:a3}

\begin{figure*}[htp]
\centering


\noindent\begin{minipage}{0.15\textwidth}
{\bf BR2}\\ $S_+, S_-$ \quad  $\longrightarrow$  \\
\end{minipage}%
\begin{minipage}{0.75\textwidth}
\centering
\includegraphics[width=.9\linewidth,height=.28\linewidth]{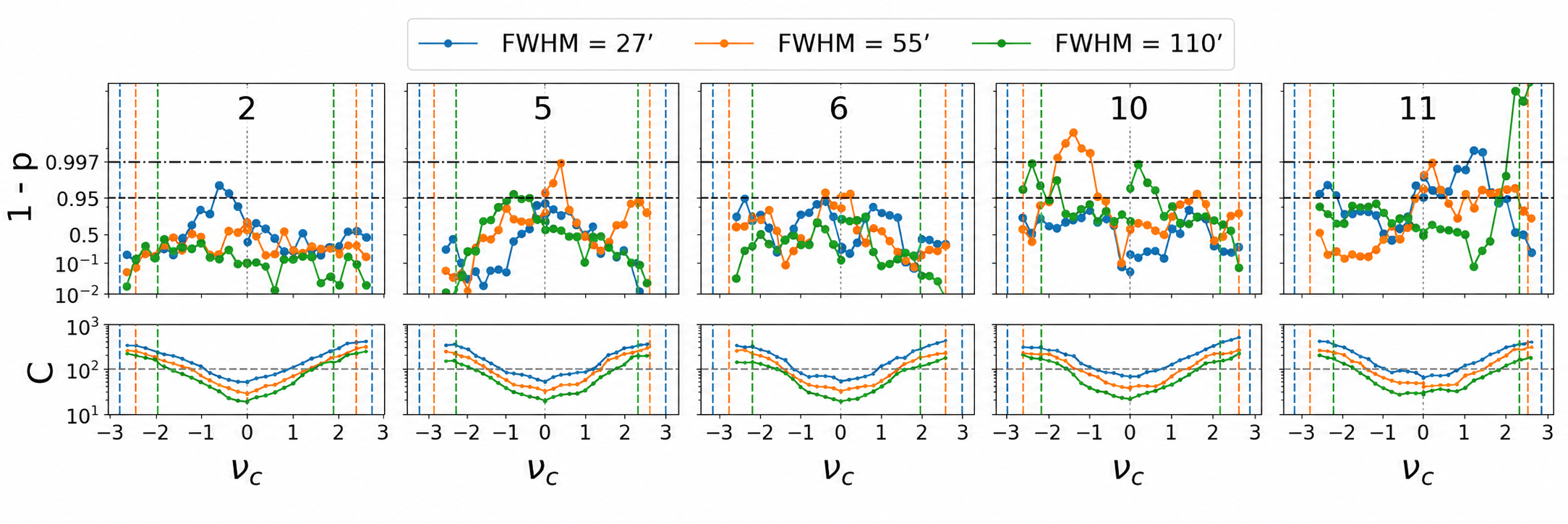}
\end{minipage}\\[0.8cm]


\noindent\begin{minipage}{0.15\textwidth}
{\bf BR4}\\ $S_+, S_-$ \quad  $\longrightarrow$  \\
\end{minipage}%
\begin{minipage}{0.75\textwidth}
\centering
\includegraphics[width=.9\linewidth,height=.76\linewidth]{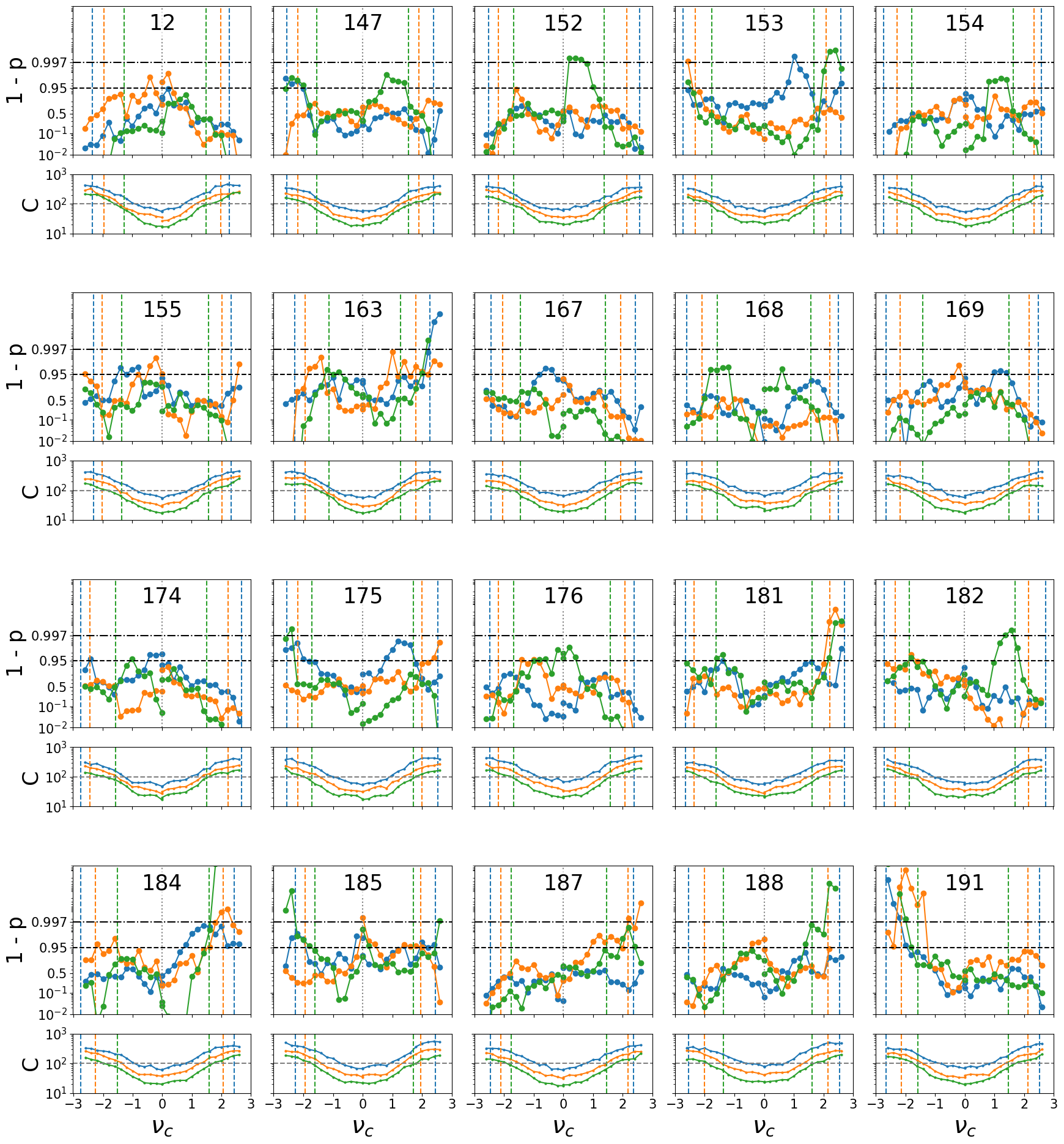}
\end{minipage}\\[0.8cm]

\noindent\begin{minipage}{0.15\textwidth}
{\bf BR4}\\ $\alpha_+,\alpha_-$ \quad  $\longrightarrow$  \\
\end{minipage}%
\begin{minipage}{0.75\textwidth}
\centering
\includegraphics[width=.4\linewidth,height=.2\linewidth]{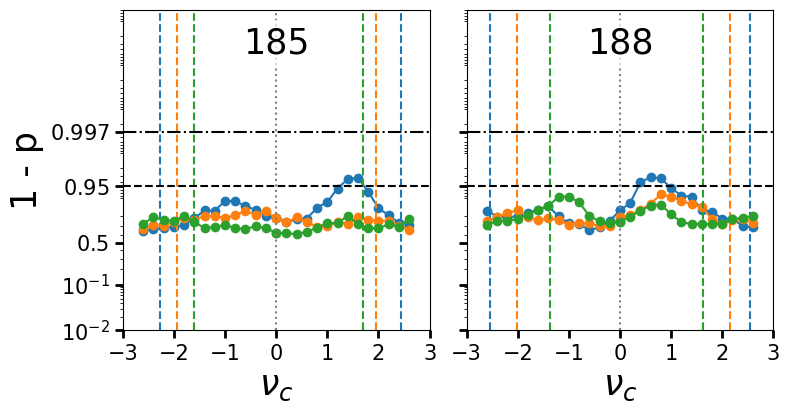}
\end{minipage}

\caption{
$(1-p)$-values versus $\nu_e$ for outlier patches (see section 5.2.2) for $S_{+}$, $S_{-}$ (both on the same panels) and $\alpha$ for BR4.
Each pair of vertical lines of each colour indicate the threshold range over which all simulations take values, for each $N_{\rm table}$.
The lower averaging threshold range is one for all plots.
The bottom sub-plots show the condition number ($C$). The gray dashed line marks $C=100$ as a visual guide.
For $\alpha$, since the covariance matrix is not used to identify anomalous patches, we do not calculate $C$.
}
\label{fig:pplots}
\end{figure*}

In figure \ref{fig:pplots}, we show the plots of $(1-p)$-values versus $\nu_c$ for outlier patches (see section \ref{sec:s5.2.2}) for $S_+,\, S_-$  (both on the same panels) and $\a$, for BR2 and BR4. Different colours indicate different FWHM values, 
as indicated. Each pair of vertical lines of each colour indicate the threshold range over which {\em all} simulations take values. The boxcar averaging threshold range is one, for all plots. These plots provide additional information of the threshold regions where the deviations are large, as well as how statistically significant the deviations are. 

A feature worth mentioning is that in figure \ref{fig:pplots}   we can notice discontinuity at $\nu_c=0$ for most of the $(1-p)$ curves. This would not be the case if we did not include Betti numbers in our analysis. Betti numbers carry additional information, particularly at intermediate thresholds $-1\lesssim \nu \lesssim 1$, compared to $V_2$. This information gets canceled out in $V_2$ as the alternating sum of the Betti numbers (neglecting the sky curvature). 

In the bottom subplots, we show the condition number ($C$) versus $\nu_c$.  Here $C$ is computed from the covariance matrix constructed over the threshold range $[\nu_c - 0.4, \nu_c + 0.4]$, centered at each $\nu_c$, and evaluated separately for $S_+$ and $S_-$ regions. This $\Sigma$ includes both auto-correlation ($\Sigma_{X_i X_i}$) and cross-correlation ($\Sigma_{X_i X_j}$) blocks as shown in figure \ref{fig:cov}. We obtain $C$ values below 100 for roughly $|\nu| \lesssim 1$, and it remains  of order 100 within the threshold ranges of interest. This indicates well-conditioned behavior for the covariance matrices.For $\a$, the $p$-value calculation does not use cross-correlations (as mentioned in Section \ref{sec:s5.2.2}), and therefore, we do not calculate the condition number $C$.

\section{Analysis of null Maps}
\label{sec:a4} 

In the context of this paper, the null map (described in section \ref{sec:sec2b}) serve to carry out a `null test' by comparing morphological statistics calculated for the observed and simulated null maps on sky patches, as done for the convergence maps in section \ref{sec:sec5b}. The premise is that if an anomalous feature appears in the null map, and a similar anomaly is seen in the observed convergence map, then, the latter is likely to originate from either foregrounds or instrument systematics. 

\begin{figure*}[htp]
\centering

\noindent\begin{minipage}{0.15\textwidth}
{\bf BR2}\\ $S_+, S_-$ \quad  $\longrightarrow$  \\
\end{minipage}%
\begin{minipage}{0.75\textwidth}
\centering
\includegraphics[width=.95\linewidth,height=.27\linewidth]{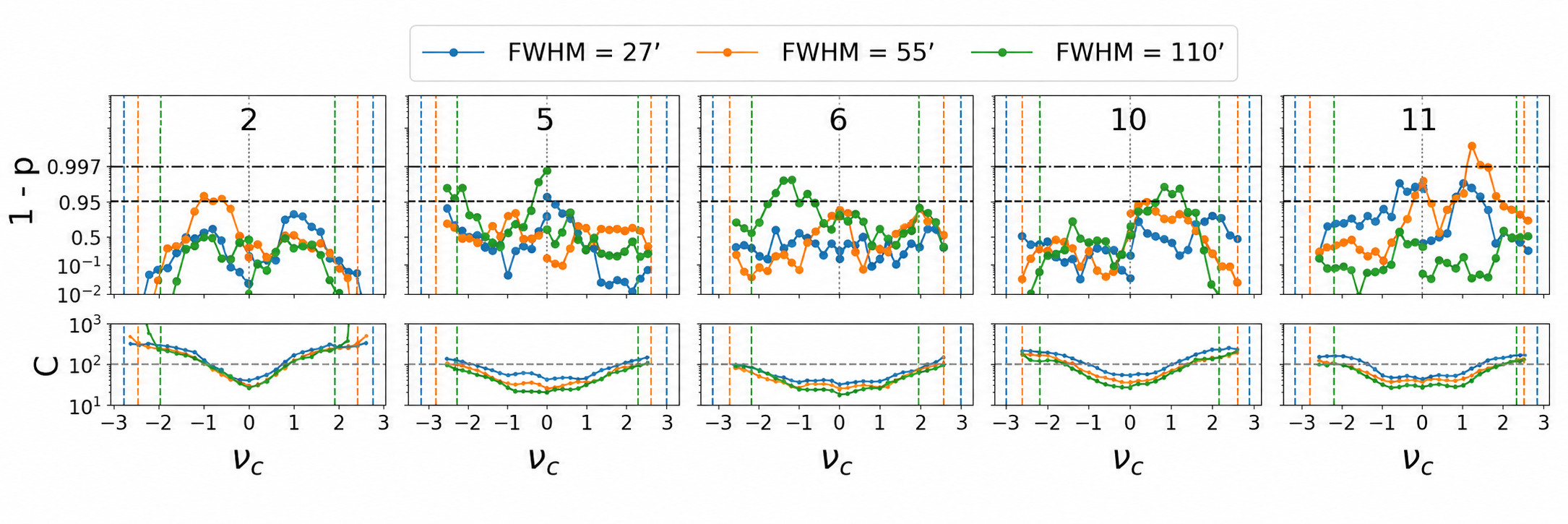}
\end{minipage}\\[0.8cm]


\noindent\begin{minipage}{0.15\textwidth}
{\bf BR4}\\ $S_+, S_-$ \quad  $\longrightarrow$  \\
\end{minipage}%
\begin{minipage}{0.75\textwidth}
\centering
\includegraphics[width=.95\linewidth,height=.6\linewidth]{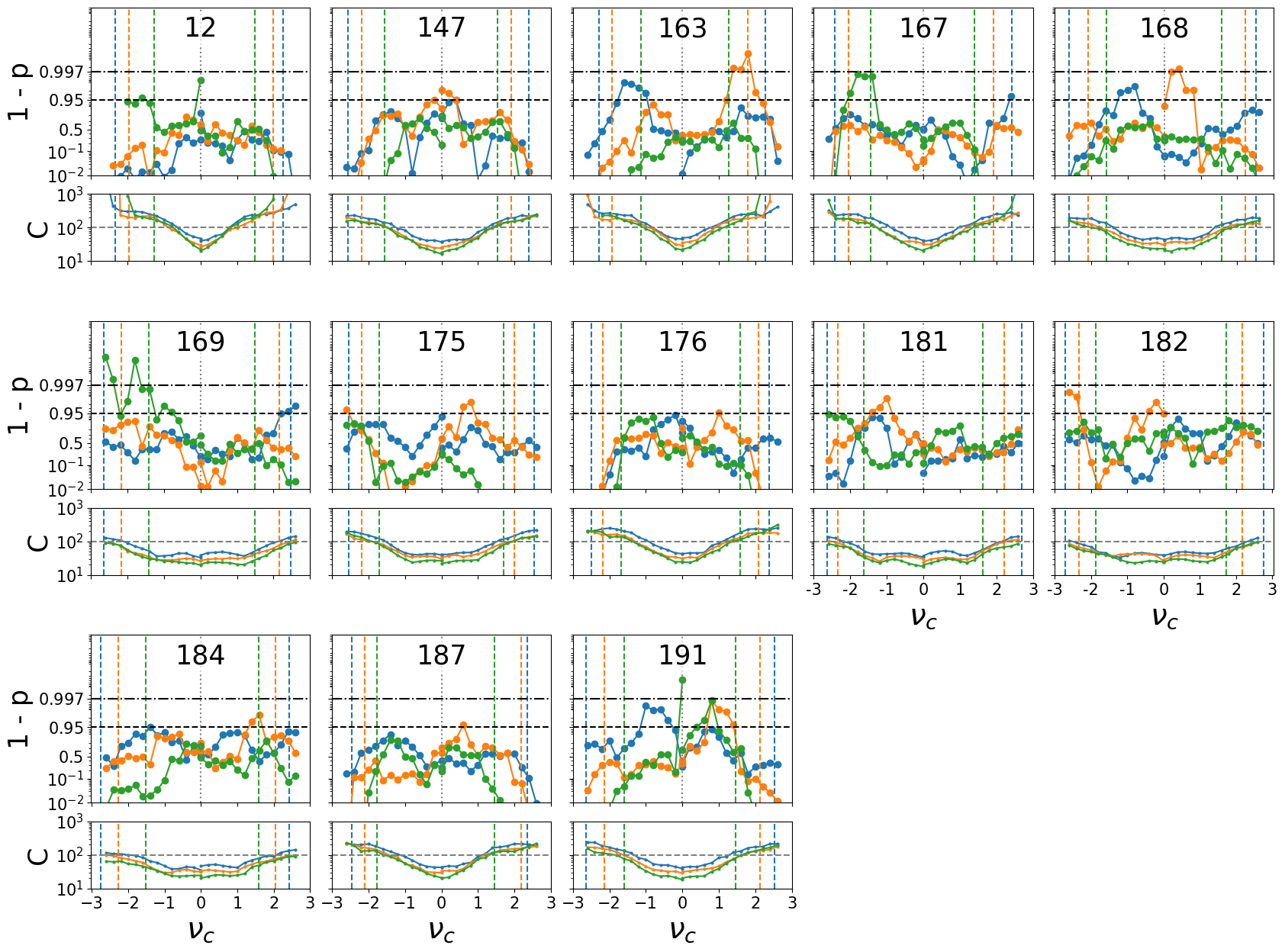}
\end{minipage}

\caption{
$(1-p)$-values versus $\nu_c$ for outlier patches (see section \ref{sec:s5.2.2}) for Null Maps (both on the same panels).
Each pair of vertical lines of each colour indicate the threshold range over which {\em all} simulations take values, for each smoothing FWHM. 
The boxcar averaging threshold range is one for all plots. The bottom sub-plots show the condition number ($C$). Again, the gray dashed line marks $C=100$ as a visual guide.}

\label{fig:pplots_null}
\end{figure*}

We compute $S_+$ and $S_-$ for the observed and simulated null maps, for each FWHM. 
For consistency with the data processing of the convergence maps in the main text, Wiener filtering (see eq. (\ref{eq:WF})) is applied to all the maps before computing the statistics.
The results are then compared between the observed and simulated null maps. 
In figure~\ref{fig:pplots_null},  we plot $(1 - p)$ as a function of $\nu_c$ for all patches identified as anomalous for the null maps in BR2 and BR4, across different smoothing 
resolutions. 
We also plot the condition number ($C$) 
in the lower sub-panels. In the case of BR2, from the results for $S_+$ statistic for both FWHM=$27'$ and $55'$,  
we find that patch 11, which was previously flagged as highly anomalous, can be contaminated by systematic artifacts, since similar deviations are also obtained for  the null maps. In the case of BR4, we find that patch 184 (for $S_+$ at FWHM=$55'$ 
and patch 191 (for $S_-$ at FWHM=$27'$ 
also show significant deviations from null expectations, suggesting that these features are likely to originate from residual systematics. The remaining patches shown in figure~\ref{fig:pplots_null} do not show up as anomalous patches for the convergence map. 

\section{Identification of anomalous patches using $\a$ statistic following the methodology of \cite{Goyal:2021}}
\label{sec:a5}

In~\cite{Goyal:2021}  the $\a$ statistic was used to identify anomalous patches in $\kappa$ map from Planck data. ACT has considerably smaller sky coverage compared to Planck which makes a direct comparison  between the two difficult. However, we can compare the overlapping regions, most of which are in the southern hemisphere.  The construction of patches is the same as done here, and FWHM values were $55'$ and $110'$. 
The criteria for anomaly were (a) higher than $2\s$ deviation at more than 2 thresholds (not necessarily consecutive), and, (b)  persistence of outlier behaviour at both FWHM 
values. The threshold bin used was $\D \nu=0.25$, whereas, here it is 0.2. This small difference does not impact the comparison. The statistical inference quantification method is similar. 

\begin{figure*}[htp]
\centering

\noindent\begin{minipage}{0.3\textwidth}
\centering 
{\bf BR4}, $\alpha$, FWHM=$27'$ \quad 
$\longrightarrow$
\end{minipage}%
\begin{minipage}{0.7\textwidth}
\centering
\includegraphics[width=.6\textwidth]{{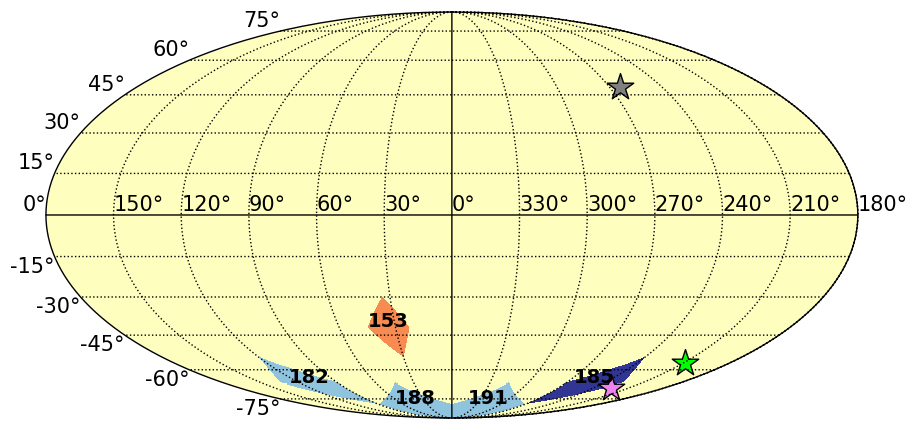}}
\end{minipage}\\[0.5cm]

\noindent\begin{minipage}{0.3\textwidth}
\centering 
{\bf BR4}, $\alpha$, FWHM=$55'$  \quad 
$\longrightarrow$
\end{minipage}%
\begin{minipage}{0.7\textwidth}
\centering
\includegraphics[width=.6\textwidth]{{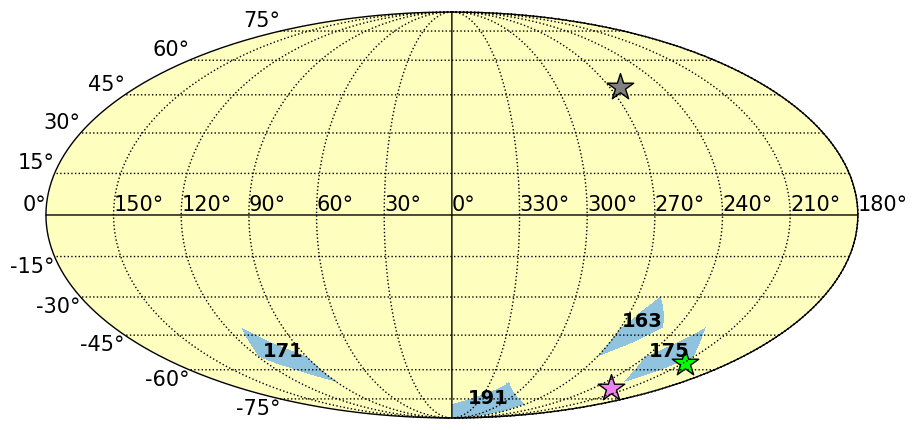}}
\end{minipage}\\[0.5cm]

\noindent\begin{minipage}{0.3\textwidth}
\centering 
{\bf BR4}, $\alpha$, FWHM=$110'$  \quad 
$\longrightarrow$
\end{minipage}%
\begin{minipage}{0.7\textwidth}
\centering
\includegraphics[width=.6\textwidth]{{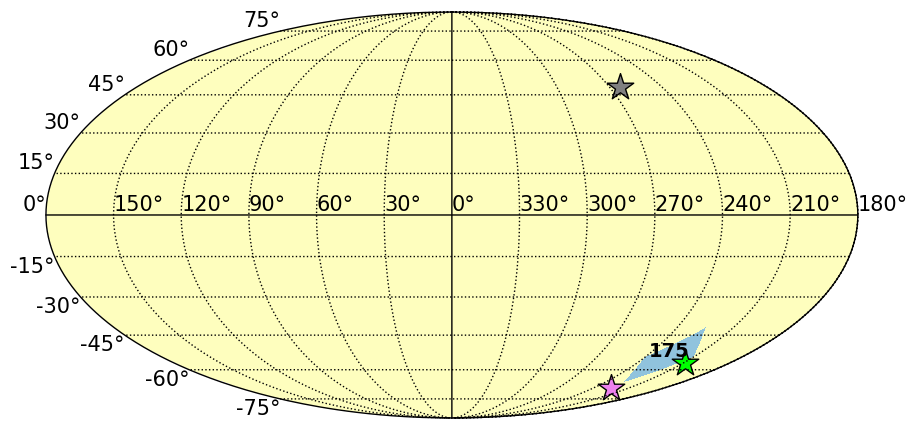}}
\end{minipage}

\caption{Anomalous patches identified using the methodology of~\cite{Goyal:2021} for FWHM=$27', 55'$ and $110'$ 
for BR4. 
This method does not distinguish positive and negative threshold. None were found for BR2.
The orange colour indicates higher than 2$\sigma$ significance, with $\alpha$ of ACT data {\em higher} than the median from simulations, 
light and dark blue indicate higher than 2$\sigma$ and $3\sigma$, respectively, with $\alpha$ of ACT data {\em lower} than the median.
The coloured stars are the same as shown in figure \ref{fig:apatches}.
}
\label{fig:priya_analysis}
\end{figure*}

To make a fair comparison with~\cite{Goyal:2021}, we first identify outlier patches using criterion (a) for the FWHM values used were - $27', 55'$ and $110'$. 
This method does not distinguish positive and negative threshold. We do not find any outliers for BR2. The outlier patches found for BR4 are shown in figure \ref{fig:priya_analysis}. The orange colour  indicates outlier at higher significance than 2$\s$, with $\a$ of ACT data found to be {\em higher} than the median simulation value, while light and dark blue  indicate higher significance than 2$\s$ and $3\s$, respectively, with $\a$ of ACT data {\em lower} than the median.  The coloured stars are the same as shown in figure \ref{fig:apatches}. Next, we see that patch 175 is common between FWHM=$55'$ and $110'$. 
Hence, using criterion (b), we identify this patch as anomalous. By comparing with figure 6 of~\cite{Goyal:2021} we conclude that there are no common anomalous patches. 
Note that patch 163 which is an outlier for FWHM=$55'$ 
is also identified as mildly anomalous in figure 6 of~\cite{Goyal:2021}, but with the difference that $\a$ of Planck data is higher than the median simulation value, while for ACT data it is lower. 

Our findings for $\a$ here suggest that some of the anomalous patches identified in the earlier work are likely due to the effect of noise in the convergence map reconstructed from Planck data.  It is also interesting to note that all anomalous patches identified here, 
except 153,  show $\a$ lower than the median simulation values, indicating a tendency towards directionally aligned structures. Patch 185 which is close to the SS shows alignment at 3$\s$. Moreover, some of these patches are the ones flagged as anomalous using $S_+$ and $S_-$ in our main results. We mention here the ones that show deviations at two or consecutive thresholds. At FWHM=$27'$, 
191 shows alignment for $\a_-$ at consecutive thresholds.  At FWHM=$55'$, 
patches 163 and 191 show alignment for $\a_-$. At FWHM=$110'$, 
patch 175 shows alignment for $\a_+$.


\def\apj{ApJ}%
\def\mnras{MNRAS}%
\def\aap{A\&A}%
\def\apjl{ApJ}
\def\aj{AJ}
\def\physrep{PhR}
\def\apjs{ApJS}
\def\jcap{JCAP}
\def\pasa{PASA}
\def\pasj{PASJ}
\def\nat{Natur}
\def\apss{Ap\&SS}
\def\araa{ARA\&A}
\def\aaps{A\&AS}
\def\ssr{Space Sci. Rev.}
\def\pasp{PASP}
\def\na{New A}
\def\prd{PRD}
\def\mathann{Math. Ann.}
\def\AAP{Adv. Appl. Prob.}

\bibliographystyle{apsrev}
\bibliography{reference}

\end{document}